\documentclass[useAMS,usenatbib,preprint]{aastex}

\usepackage{times}
\usepackage{graphicx}
\usepackage{xcolor}
\usepackage{array}
\usepackage{amssymb}
\usepackage{amsmath}
\usepackage{natbib}
\usepackage{booktabs}
\usepackage{multirow}

\title{Knot a Bad Idea:\\Testing BLISS Mapping for Spitzer Space Telescope Photometry}
\author{J. C. Schwartz$^{1,2,3}$ \thanks{E-mail: joelschwartz2011@u.northwestern.edu} \ \thanks{McGill Space Institute (McGill U.); Institute for Research on Exoplanets (UdeM)} \ \& N. B. Cowan$^{2,3 \ \dagger}$\\
\footnotesize$^{1}$Department of Physics \& Astronomy, Northwestern University, 2145 Sheridan Road, Evanston, IL, 60208, USA\\
\footnotesize$^{2}$Department of Earth \& Planetary Sciences, McGill University, 3450 rue University, Montreal, QC, H3A 0E8, CAN\\
\footnotesize$^{3}$Department of Physics, McGill University, 3600 rue University, Montreal, QC, H3A 2T8, CAN\\
\footnotesize Accepted in PASP: September 26th, 2016}

\begin{document}
	
\maketitle

\section*{Abstract}
Much of transiting exoplanet science relies on high-precision photometry. The current generation of instruments can exhibit sensitivity variations greater than the astrophysical signals. For the InfraRed Array Camera (IRAC) on the Spitzer Space Telescope, a popular way to handle this is BiLinearly-Interpolated Subpixel Sensitivity (BLISS) mapping. As part of a Markov Chain Monte Carlo (MCMC), BLISS mapping estimates the sensitivity at many locations (knots) on the pixel, then interpolates to the target star's centroids. We show that such embedded optimization schemes can misfit or bias parameters. Thus, we construct a model of \emph{Spitzer} eclipse light curves to test the accuracy and precision of BLISS mapping. We compare standard BLISS mapping to a variant where the knots are fit during the MCMC, as well as to a polynomial model. Both types of BLISS mapping give similar eclipse depths, and we find that standard knots behave like real parameters. Standard BLISS mapping is therefore a reasonable shortcut to fitting for knots in an MCMC. BLISS maps become inaccurate when the photon noise is low, but typically approximate the real sensitivity well. We also find there is no perfect method for choosing the ideal number of BLISS knots to use on given data. BLISS mapping gives fits that are usually more accurate than precise (i.e. they are overly conservative), and the routine is more precise than polynomial models for significant eclipses or pixels with more varied sensitivities. BLISS mapping has better predictive power for most of these particular synthetic data, depending on how one treats time-correlated residuals. Overall, we conclude that BLISS mapping can be a reasonable sensitivity model for IRAC photometry.

\emph{Keywords:} eclipses --- instrumentation: detectors --- methods: data analysis --- methods: statistical

\section{Introduction}
\label{sec:intro_knot}
It is hard to characterize the atmospheres of transiting exoplanets because the atmospheric signal is $10^{-3}$--$10^{-5}$ of the stellar flux \citep{seager2010exoplanet}. Unfortunately, most current telescopes and instruments were not designed for these precisions.

Consider the Spitzer Space Telescope \citep{werner2004spitzer}: many planets have been observed with its InfraRed Array Camera \citep[IRAC;][]{fazio2004infrared}, and these light curves are a large part of the available data \citep[e.g.][]{Agol_2010,nymeyer2011spitzer,mahtani2013warm,wong2015W14b}. The pixels in IRAC are not uniformly sensitive and the target centroid (i.e. stellar position) moves on timescales of minutes to days \citep{ingalls2016repeatability}. That means IRAC can distort the light we see \citep[e.g.][]{crossfield2012spitzer}.

Many detector models have been used to deal with sensitivity variations on a pixel. Early analyses of \emph{Spitzer} light curves used polynomials \citep{charbonneau2005detection,knutson20083}. \citet{ballard2010search,ballard2011kepler} used Kernel Regression to analyze IRAC and Kepler Space Telescope data; improved versions of this method were used by \citet{knutson20123}, \citet{lewis2013orbital}, and \citet{wong2015W14b,wong20163}. \citet{morello2014new} used Independent Component Analysis \citep[ICA;][]{waldmann2012cocktail} to reanalyze IRAC transit light curves. More recently, \citet{deming2015spitzer} used Pixel-Level Decorrelation (PLD) to remove red noise from IRAC data. The authors state this method is better than modeling the sensitivity with centroids for a few reasons, including that PLD is analytically sound and runs fast.

In recent years, many researchers have used BiLinearly-Interpolated Subpixel Sensitivity mapping \citep[BLISS hereafter;][]{stevenson2012transit}. This routine works quickly in a Markov Chain Monte Carlo (MCMC) because no \emph{explicit} parameters are used for the detector sensitivity. Instead, BLISS divides the light curve by the current astrophysical signal at each MCMC step, averages the leftover residuals at many locations on the pixel (``knots"), then interpolates to find the sensitivity at each centroid. This means BLISS optimizes the sensitivity at each knot---it runs efficiently because the weight of each knot at the centroids' locations can be calculated ahead of time.

Many studies have used BLISS to model the intra-pixel sensitivity in \emph{Spitzer} data, as shown in Table \ref{tab:BLISS_planets}. \citet{lanotte2014global} and \citet{demory2016variability,demory2016map} also included the full-width half-maximum of the pixel response function in their analyses. A recent study by \citet{ingalls2016repeatability} found that BLISS, PLD, and ICA are the most accurate and reliable ways to model IRAC sensitivity for real and synthetic observations of XO-3b. These methods can usually fit eclipse depths to within $3\times$ the photon limit of the true values.

\begin{table}[h!]
	\centering
	\caption[Works that model \emph{Spitzer} intra-pixel sensitivity with BLISS.]{Works that use BLISS to model the intra-pixel sensitivity in \emph{Spitzer} IRAC data.}
	\footnotesize
	\label{tab:BLISS_planets}
	\begin{tabular}{l l}
		\toprule
		\multicolumn{1}{c}{\bfseries Reference} & \bfseries Planet/System \\
		\midrule
		\citet{stevenson2012transit} & HD 149026b\\
		\citet{stevenson2012two} & GJ 436\\
		\citet{lanotte2014global} & ...\\
		\citet{blecic2013thermal} & WASP-14b\\
		\citet{cubillos2013wasp} & WASP-8b\\
		\citet{blecic2014spitzer} & WASP-43b\\
		\citet{cubillos2014spitzer} & TrES-1\\
		\citet{diamondlowe2014new} & HD 209458b\\
		\citet{gillon2014search} & GJ 1214\\
		\citet{stevenson2014deciphering} & WASP-12b\\
		\citet{stevenson2014transmission} & ...\\
		\citet{motalebi2015harps} & HD 219134b\\
		\citet{triaud2015wasp} & WASP-80b\\
		\citet{yu2015tests} & PTFO 8-8695 b\\
		\citet{demory2016variability} & 55 Cnc e\\
		\citet{demory2016map} & ...\\
		\citet{stevenson2016search} & HAT-P-26b\\
		\bottomrule
	\end{tabular}
\end{table}

However, BLISS does not fit for the detector sensitivity---it merely optimizes it. The BLISS maps vary during an MCMC, but they always do so jointly with the astrophysical model. Thus, one cannot explore the full parameter space because the BLISS map and astrophysical model are not chosen independently (Section \ref{sec:marg_opt}). With large numbers of BLISS knots, one can also end up fitting noise in the light curve. Both of these issues mean BLISS may give astrophysical uncertainties that are too small \citep{hansen2014broadband}.

BLISS was introduced to side-step the computational challenge of a fully Bayesian approach \citep{stevenson2012transit}. However, nobody has tested the impact of this shortcut, nor has anybody published a rigorous study of BLISS using synthetic \emph{Spitzer} observations, for which one knows the ground truth. \citet{ingalls2016repeatability} tested seven techniques for removing correlated noise from IRAC data using real and synthetic observations---but only for a single hot Jupiter, XO-3b. We will therefore investigate BLISS by using a simple model of \emph{Spitzer} IRAC light curves.

\citet{stevenson2012transit} created BLISS to handle the intra-pixel sensitivity in IRAC data because fitting $\sim10^{5}$ measurements with $\sim10^{3}$ model parameters in an MCMC was not feasible. This is still true, so we test light curves that have a modest number of data by using $\sim25$--$150$ BLISS knots (but see Sections \ref{sec:choose_mesh} and \ref{sec:more_synth_fits}). These sets of parameters are small enough that we can \emph{directly} fit each knot.

We organize our work as follows: in Section \ref{sec:marg_opt}, we describe how properly marginalizing a parameter differs from optimizing it, and use examples to show that this can affect the fits on other parameters. Then, in Sections \ref{sec:toy_model} and \ref{sec:toy_fits}, we use a toy model to show that optimizing may cause problems even with simple posteriors and Gaussian uncertainties. We describe our model of the \emph{Spitzer} IRAC detector in Section \ref{sec:D_model}, including how we make mock centroids, then introduce our astrophysical model and synthetic light curves in Section \ref{sec:A_model}. In Section \ref{sec:B_routine}, we briefly review BLISS, and in Section \ref{sec:map_compare}, we compare BLISS knots and maps to the true pixel sensitivity. We then fit our light curves with MCMC and three different models for the pixel sensitivity, including two versions of BLISS, in Section \ref{sec:synthetic_fits}. We discuss our results in Section \ref{sec:discuss_knot} and summarize our work in Section \ref{sec:conclude_knot}. For those interested, the details about how we choose parameters for the pixel's sensitivity and the astrophysical signal are given in Appendices \ref{sec:D_sens_explain} and \ref{sec:A_param_explain}, respectively.

\section{Optimizing Nuisance Parameters}
\label{sec:nuisance}
Nuisance parameters are parts of a study that are not interesting, but have to be used to get a good answer. In the context of characterizing transiting planets, the detector sensitivity is usually modeled in terms of nuisance parameters.

\subsection{Marginalizing vs. Optimizing}
\label{sec:marg_opt}
\begin{figure}[h!]
\centering
\includegraphics[width=0.9\linewidth]{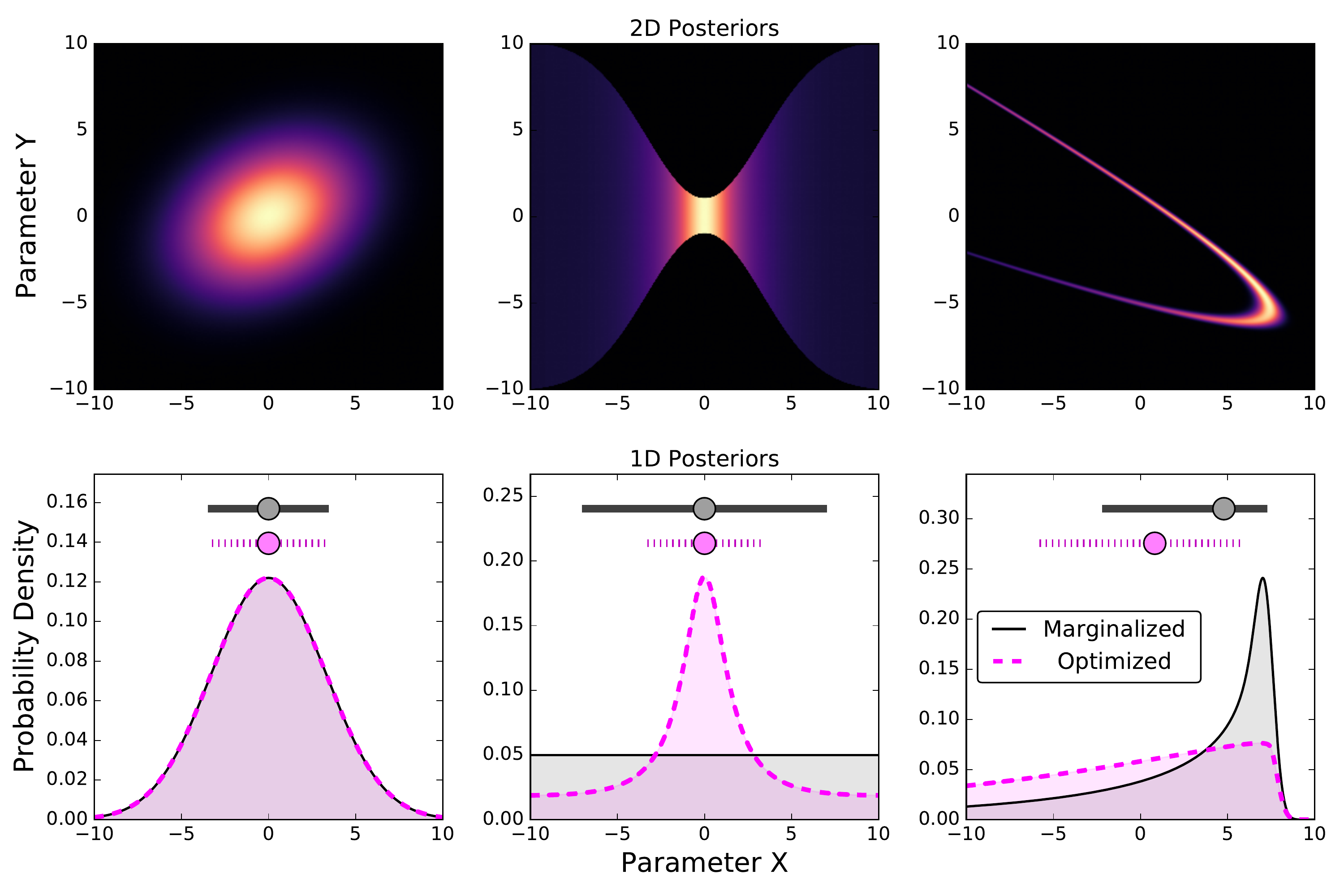}
\caption[Marginalization versus optimization for different 2D posteriors.]{\emph{Upper Panels:} Example 2D posteriors for the parameters X and Y: a bivariate Gaussian on the left, a ``Gaussian butterfly" in the center, and a Rosenbrock banana on the right. Each color scale ranges from the maximum of the posterior (light) down to zero (dark). \emph{Lower Panels:} The normalized 1D posteriors for each parameter X, where the black curves are the densities after marginalizing (i.e. integrating over or directly fitting) each parameter Y. Instead, one could optimize Y (i.e. find the most probable Y for each X) to get the densities shown by the dashed magenta curves. Slicing Y (i.e. cutting along the Y-value of the 2D peak) is not shown, but can be much different from optimizing Y (e.g. Rosenbrock banana), especially for high-dimensional posteriors. The color-coded circles are median values of each posterior, and the bars show $1\sigma$ intervals. For the bivariate Gaussian, one infers the same posterior and fit interval for X by marginalizing or optimizing Y---this does not happen in the other two cases. Optimizing a nuisance parameter can make the fit on another variable too precise, too conservative, or even biased.}
\label{fig:OptMarg_Posts}
\end{figure}

When fitting a model to data, one explores a posterior probability function: this describes how likely one's model is given each choice of parameter values. Posteriors often have many dimensions, so we show a bivariate Gaussian as a simplified example in the upper left panel of Figure \ref{fig:OptMarg_Posts}. This posterior describes the arbitrary parameters X and Y, where the lighter colors show pairs of parameters that are more probable. Even though this 2D Gaussian is not oriented along X or Y, it is still highly symmetric.

Suppose now that parameter Y is a nuisance variable, and one would like the posterior (i.e. the fit) for the ``interesting" parameter X alone. There are three general ways to find this, though we will focus on two for the moment. Ideally one should marginalize over Y, or integrate the 2D posterior over all possible Y-values, as shown by the (normalized) black curve in the lower left panel of Figure \ref{fig:OptMarg_Posts}. Instead one could try optimizing Y, or finding the highest probability along Y for each value of X, shown in the same panel as a dashed magenta curve. For the bivariate Gaussian both methods give identical 1D posteriors on X: the median of each curve is shown with a color-coded circle, while the bars are the $1\sigma$ intervals. In other words, how one deals with this nuisance parameter Y does not affect their fit for X.

%\begin{sidewaysfigure}
%\end{sidewaysfigure}

Some posteriors are less well-behaved; we show two examples in the remaining panels of Figure \ref{fig:OptMarg_Posts}. The 2D posterior in the upper center is a ``Gaussian butterfly," which has a narrow range of defined Y-values around $\mathrm{X}=0$ that broadens as $|\mathrm{X}|$ increases. The probability density varies only along X and is inversely related to the width in Y---that means the marginalized posterior for X is flat (black curve in the lower center) and the optimized version peaks at $\mathrm{X}=0$ (dashed magenta curve). If one optimizes this parameter Y, their median value for X is correct (circles) but their uncertainty is too small (bars).

Alternatively, consider a 2D posterior shaped like a Rosenbrock banana function in the upper right panel of Figure \ref{fig:OptMarg_Posts}. This has two thin branches that join near $(\mathrm{X},\mathrm{Y})=(7,-5)$, and the probability density does not vary the same way in both branches. The posterior for parameter X after marginalizing Y, in the lower right panel, is denser on the right and peaks around $\mathrm{X}=7$. By optimizing Y, though, one misses most of the banana's lower branch and so gets a flatter 1D posterior on X. In this case, the uncertainty on X is \emph{larger} when optimizing Y, and the median is biased towards smaller X-values.

The third method we alluded to for fitting parameter X is slicing the given 2D posterior along the Y-value at its peak. This is nearly the same as optimizing Y for our first two examples, but with the Rosenbrock banana the 1D posterior for X has just two narrow, distinct peaks (not shown). For higher dimensional cases, optimizing typically falls somewhere between marginalizing and slicing the full posterior. We will return to this idea when testing BLISS in an MCMC in Section \ref{sec:fiduc_BJ}.

In general, then, optimizing parameters works well when it approximates marginalizing over those parameters: having just the silhouette of the posterior seen by the interesting variable(s) is enough to describe the nuisance parameter(s) throughout the space. This is true for the bivariate Gaussian, and in principle for multivariate Gaussians, too. Once the posterior is non-convex, has an exotic density profile, or is otherwise oddly shaped, optimizing along one or more dimensions is dicey. This may bias the best-fit values of interesting parameters and make it hard to report reasonable uncertainties.

\subsection{Toy Model}
\label{sec:toy_model}
\begin{figure}[h!]
	\centering
	\includegraphics[width=0.9\linewidth]{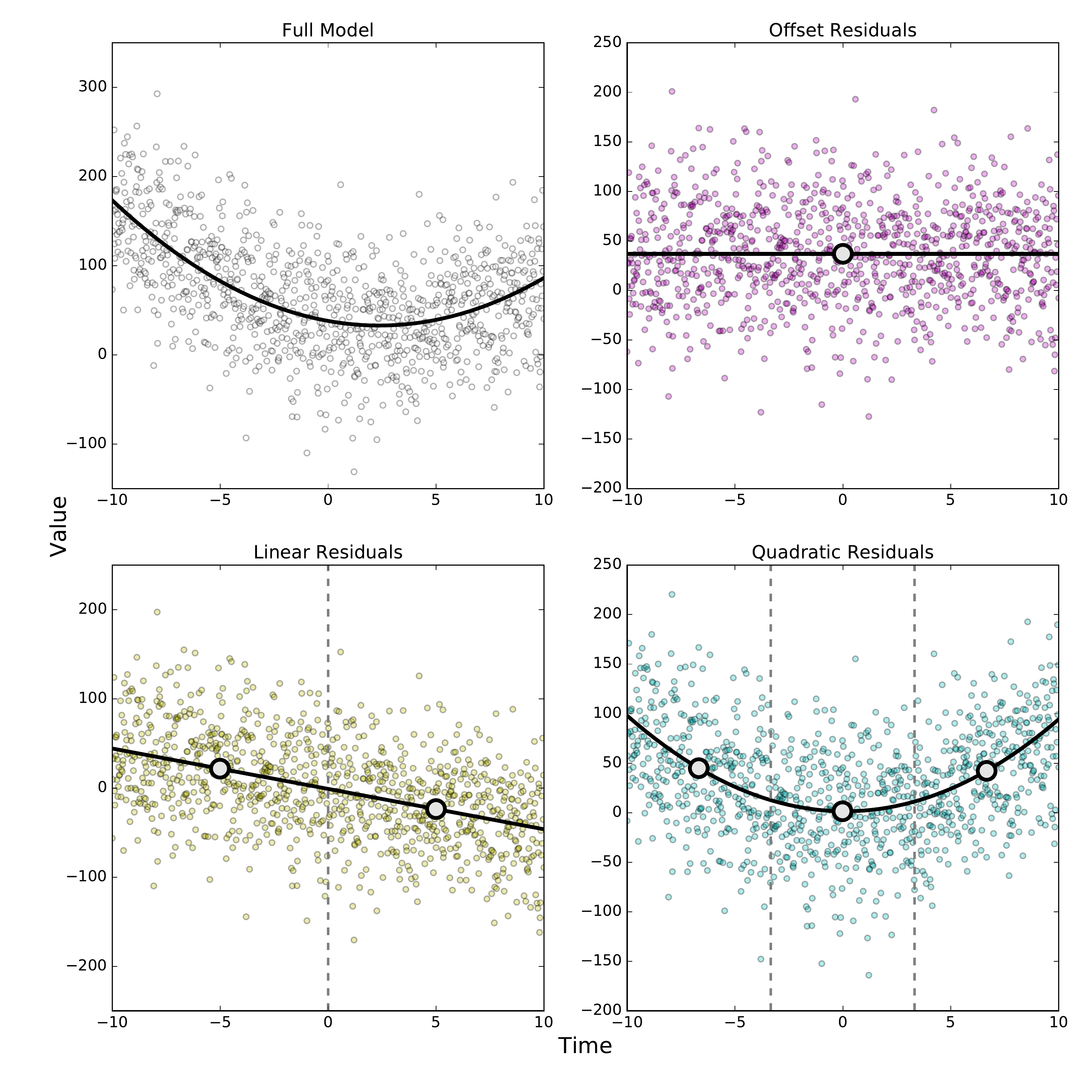}
	\caption[Example data and residuals from a toy polynomial model.]{\emph{Upper Left:} Example data generated from Equation \ref{eq:toy_2nd_poly}, where the black curve is the true function without noise. \emph{Other Panels:} Residuals left after subtracting three incomplete models, with no offset, linear, or quadratic term, from the data at upper left: $b$-Optimize at upper right (magenta), $m$-Optimize at lower left (yellow), and $q$-Optimize at lower right (cyan), respectively. One can estimate each missing term by splitting the residuals into time groups (dashed vertical lines), finding the mean (large gray circles) of each group, and getting the leading part of the trend (black curves) through these means. Thus, one can try to optimize each term using data residuals.}
	\label{fig:samp_toy_poly_resids}
\end{figure}

Even if a posterior seems well-behaved, optimizing nuisance parameters can still cause problems. We demonstrate this with a toy example:
\begin{eqnarray}
\label{eq:toy_2nd_poly}
f(t) = (qt^{2} + mt + b) + N(t;\sigma),
\end{eqnarray}
where $f(t)$ is data at time $t$, the $q$, $m$, and $b$ are coefficients, and $N(t;\sigma)$ is Gaussian noise with uncertainty $\sigma$. A sample data set from this toy model is shown in the upper left panel of Figure \ref{fig:samp_toy_poly_resids}. We use 1001 evenly-spaced times, $t \in [-10,10]$, for a chosen set of parameters, $\lbrace q,m,b,\sigma \rbrace$.
%using the parameters $q=0.917$, $m=-4.347$, $b=38.046$, and $\sigma=52.288$.

The simplest way to fit these data is to use Equation \ref{eq:toy_2nd_poly}, where all four parameters are fit directly. Suppose, though, that one wanted to optimize $b$, $m$, or $q$ instead; we show examples of this strategy in the other panels of Figure \ref{fig:samp_toy_poly_resids}. This is essentially how BLISS treats pixel sensitivity \citep{stevenson2012transit}, where detector parameters are optimized and astrophysical parameters are fitted. The idea here is to make a model with the interesting variables, then subtract this incomplete model from the data to get residuals. Then one splits the residuals into groups by time, takes the mean of each group, and finds the trend through those means. As shown, this optimizes either the offset ($b$), slope ($m$), or quadratic term ($q$), described ideally in Section \ref{sec:marg_opt}. We use obvious names for each method: $b$-Optimize (upper right, magenta), $m$-Optimize (lower left, yellow), and $q$-Optimize (lower right, cyan).

\subsection{MCMC Fits to Toy Models}
\label{sec:toy_fits}
\begin{figure}[h!]
	\centering
	\includegraphics[width=0.9\linewidth]{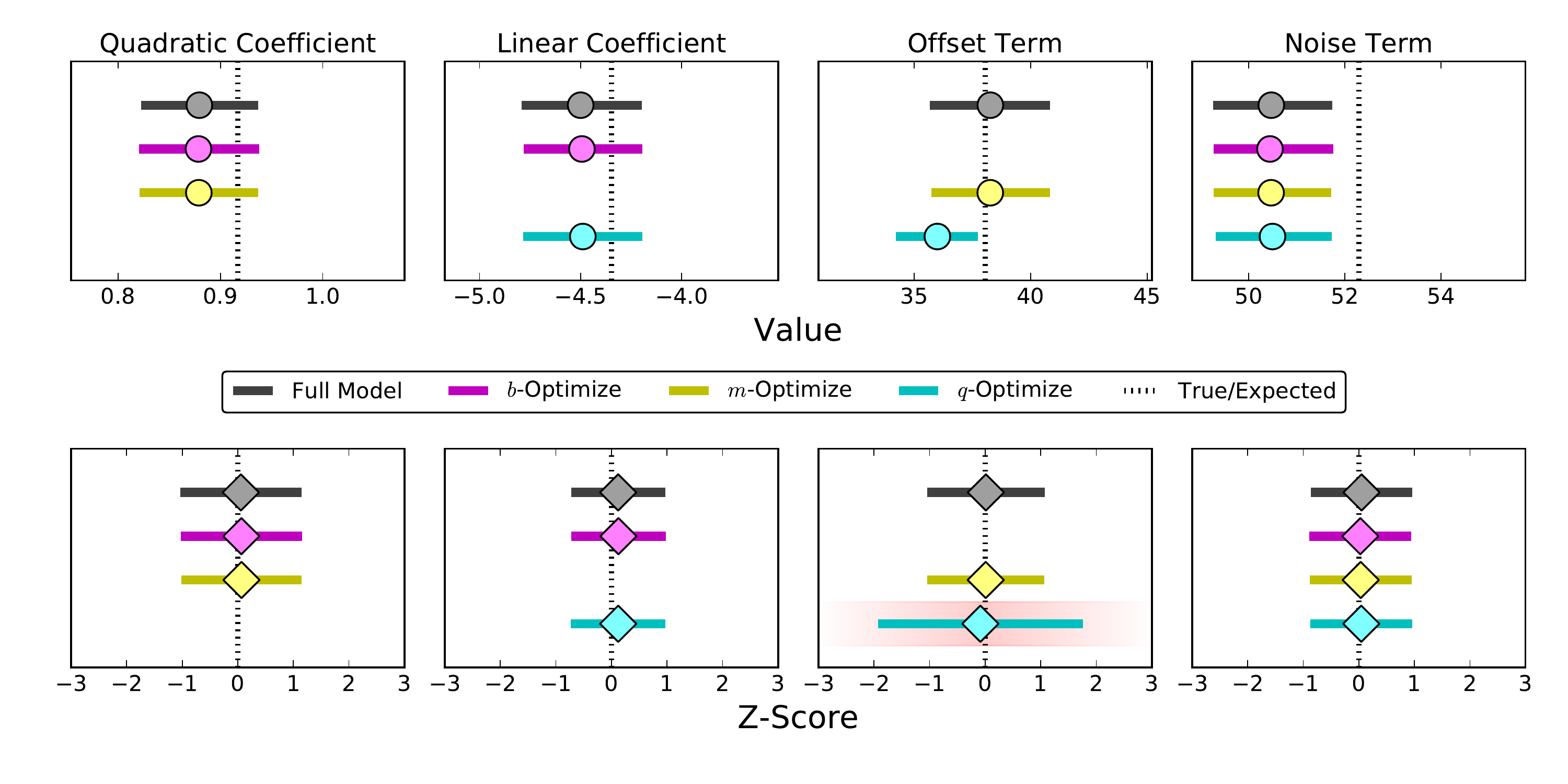}
	\caption[Example MCMC fits to the data in Figure \ref{fig:samp_toy_poly_resids}.]{\emph{Upper Panels:} Example fits to the data from the upper left panel of Figure \ref{fig:samp_toy_poly_resids}, using each of the four models described in Section \ref{sec:toy_model}. The circles are best-fit values, bars are $1\sigma$ intervals, and the dotted black lines show the true values of each parameter. \emph{Lower Panels:} Distribution of z-scores (Equation \ref{eq:z_score}) for MCMC fits to 100 random data sets from Equation \ref{eq:toy_2nd_poly}. The diamonds are mean values and the bars show $1\sigma$ intervals. The fit on a model term (e.g. upper panels) is unbiased if that term's z-scores are centered on zero---the uncertainty is reliable if they have a width of unity. The z-scores for the offset term in the $q$-Optimize model are more spread out by a factor of 2 (highlighted in red). Even with simple models and well-behaved data, optimizing a nuisance term can still lead to poor fits on interesting parameters.}
	\label{fig:samp_toy_poly_allMCMC}
\end{figure}

We now use the MCMC code \texttt{emcee} \citep{foreman2013emcee} to fit the data from Figure \ref{fig:samp_toy_poly_resids}. For each of our four models, we use 240 walkers and start them in a small ball near the true parameters. We also pick uniform priors on each term in Equation \ref{eq:toy_2nd_poly}. We burn-in each chain for 250 steps and run them for another 1000 steps, then thin the chains by the longest autocorrelation time, $\tau_{\mathrm{max}}$, that \texttt{emcee} estimates ($\tau_{\mathrm{max}} \approx 25$--$60$ steps). Example fits are shown in the upper row of Figure \ref{fig:samp_toy_poly_allMCMC}. The circles are medians of each chain and bars are $1\sigma$ intervals, as in Figure \ref{fig:OptMarg_Posts}.

%\begin{sidewaysfigure}
%\end{sidewaysfigure}

Most of the fits to the mock data are reasonable. This is no surprise for the full model---after all, we used the same four parameters to generate the data. It is also clear that one could optimize $b$ or $m$ during the MCMC without hurting anything, although these schemes run no faster than the full model.

The $q$-Optimize method is different, though. The linear and noise terms are about the same as the other three methods, but the uncertainty on $b$ is noticeably smaller. The center of the interval is also lower than the other methods. These walkers overlapped the same part of parameter space but tried a smaller range of offset terms.

We next try fitting 100 different data sets, where we randomly pick $q \in \left[-1,1\right]$, $m \in \left[-10,10\right]$, $b \in \left[-100,100\right]$, and $\sigma$ from a Normal distribution with mean 50 and width 10. We use all four methods with the same MCMC setup as before, and calculate the z-scores for each term:
\begin{equation}
\label{eq:z_score}
z_{\mu} = \frac{\mu_{\theta} - \theta}{\sigma_{\theta}},
\end{equation}
where $z_{\mu}$ is the z-score, and $\lbrace \mu_{\theta},\sigma_{\theta} \rbrace$ are the fitted value and uncertainty of parameter $\theta$. If a parameter estimate is unbiased and accurate, then the average z-score should be close to zero and the standard deviation should be close to unity. We show the z-scores in the bottom panels of Figure \ref{fig:samp_toy_poly_allMCMC}, where diamonds are the mean values.

The trend in these z-scores is obvious. As we expect, the full model, $b$-Optimize, and $m$-Optimize fits look fine: on average we get close to the real parameters and have reasonable uncertainties. This is even true for parts of $q$-Optimize, but not the offset that this method finds. In general this uncertainty on $b$ is too small, which is why the z-scores are more spread out than any other fit, by about a factor of 2. In other words, if one were to model this kind of data using $q$-Optimize, they would be too precise on their guess for $b$. Although this case mimics the \emph{fits} in the lower center panel of Figure \ref{fig:OptMarg_Posts}, the posterior for our toy model looks like a 4D ellipsoid (i.e. Go stone). Either $q$-Optimize does not ``optimize" in the sense of Section \ref{sec:marg_opt}---possible but unlikely---or the density of this posterior varies in an unexpected way.

It may seem silly to optimize the quadratic term in a quadratic equation---if one expects this term, then they should probably fit for it directly. BLISS, however, uses the same strategy to optimize the \emph{entire} detector signal, not just one part of it. As acknowledged by \citet{stevenson2012transit}, this is an expedient shortcut since fitting for $\sim10^{3}$ knot values is not computationally feasible. Our example posteriors and toy model demonstrate that this shortcut may come at the price of accurate astrophysical parameters.

\section{Synthetic Light Curves}
\label{sec:mock_LCs}

\subsection{Detector Model}
\label{sec:D_model}
We begin by simulating the \emph{Spitzer} detector. Each wavelength channel of IRAC has an array of pixels, and due to the peak-up, the centroids usually stay within a single pixel for an entire eclipse observation \citep{ingalls2016repeatability}. In real IRAC data, the image falls on different parts of the pixel because \emph{Spitzer} both shakes and drifts slightly \emph{and} has changes in optics due to thermal expansion and contraction.

\begin{figure}[h!]
	\centering
	\includegraphics[width=0.9\linewidth]{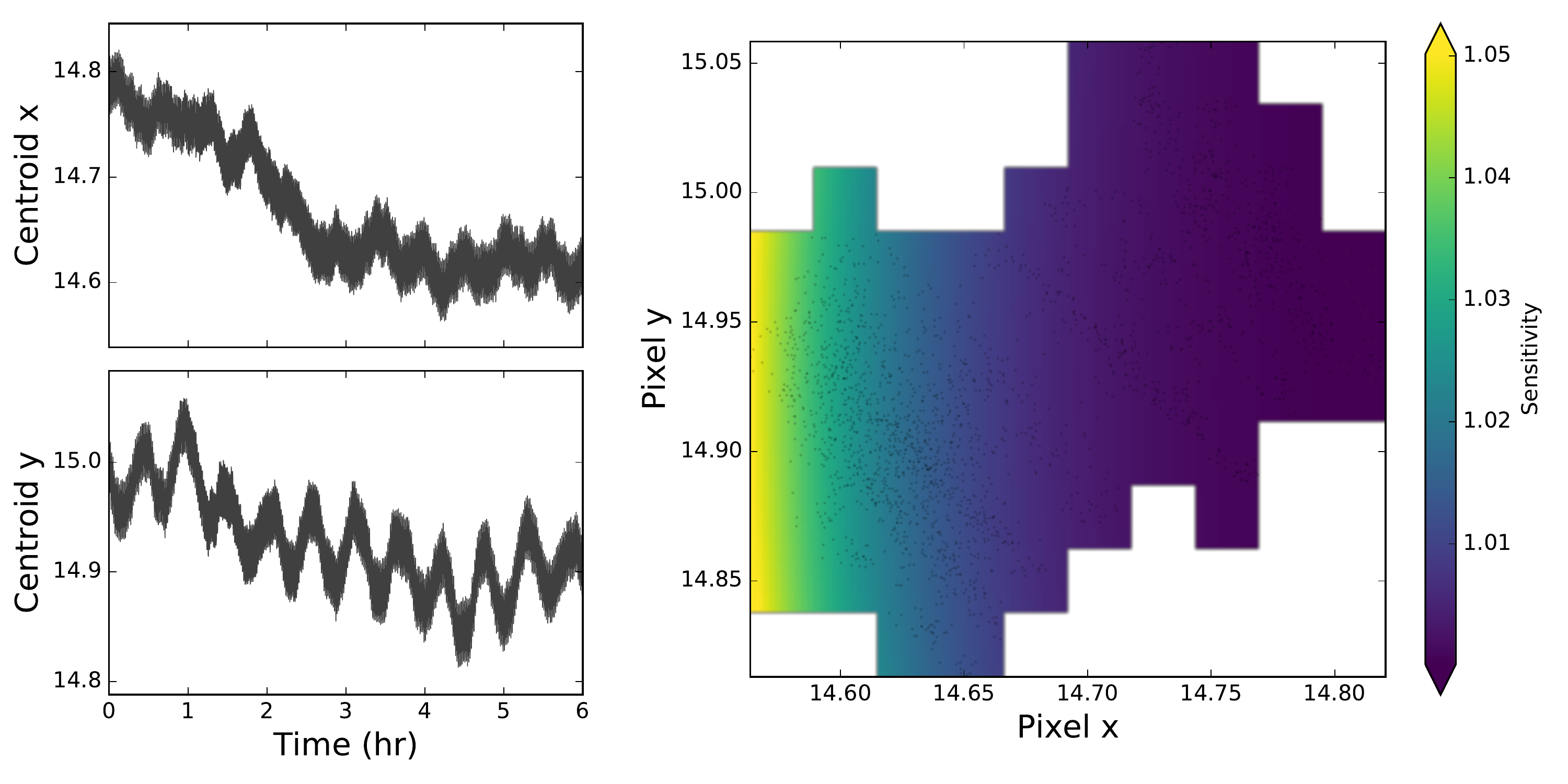}
	\caption[Example centroid positions and a pixel sensitivity map.]{\emph{Left Panels:} Traces of centroid position for a mock observation of a planetary eclipse. This observation is 6 hours long with $N=2160$ measurements, or about 10 seconds per datum. \emph{Right:} An example sensitivity map for the region of the pixel sampled by these centroids (gray dots), where lighter colors are more sensitive areas. The darkest and lightest colors are outside the sensitivity range shown---no centroids are located in these spots. We test a variety of sensitivity variations in Sections \ref{sec:map_compare} and \ref{sec:more_synth_fits}}
	\label{fig:samp_cents_pixsens}
\end{figure}

We mimic this by modeling the centroid time-series, $\lbrace x_{0}(t),y_{0}(t) \rbrace$, with the pointing equations in Appendix A1 of \citet{ingalls2016repeatability}, but make two changes. We drop their short-term drift because we assume the eclipses we will model do not happen just after a re-pointing. For full-orbit phase curves where the centroids often cover larger regions of the pixel \citep[e.g.][]{cowan2012thermalW,wong20163}, including this drift could make polynomial models (Section \ref{sec:synthetic_fits}) less accurate at describing the sensitivity variations. We also use regular, as opposed to fractional, Brownian motion to make the noise for their ``jitter" term. This change should not influence the centroids on timescales longer than 60 seconds, i.e. the jitter period. Examples of these centroids are shown in the left panels of Figure \ref{fig:samp_cents_pixsens}---this observation lasts 6 hours and has 2160 data, $N$, or about 10 seconds per point.

%\begin{sidewaysfigure}
%\end{sidewaysfigure}

The first $(x_{0},y_{0})$ are both randomly chosen from $[14.7,15.3]$ because $(x,y)$ on the central pixel both span $[14.5,15.5]$. We model this pixel's sensitivity using a polynomial:
\begin{equation}
\label{eq:sens_vars}
V(x,y) = 1 + \left( \sum_{\ell = 0}^{n} \sum_{m = 0}^{n - \ell} c_{\ell m} (x - 15)^{\ell} (y - 15)^{m} \right)_{\ell m \neq 00},
\end{equation}
where $V(x,y)$ is the sensitivity map and $n$ is the polynomial order (we use $n=7$). The $c_{\ell m}$ are coefficients, and the details about how we pick these are given in Appendix \ref{sec:D_sens_explain}. This equation keeps the average sensitivity close to unity; we show an example map in the right panel of Figure \ref{fig:samp_cents_pixsens}. The center of the pixel, $(x,y)\approx(15,15)$, is the most sensitive region on the real IRAC detector \citep[e.g.][]{reach2005absolute,cowan2012thermalW}---this is not always true for Equation \ref{eq:sens_vars}.

With the centroids and sensitivity map, we then make a detector signal, $D(t)$, using:
\begin{equation}
\label{eq:D_signal}
D(t) = V(x_{0}(t),y_{0}(t)),
\end{equation}
that has a given amplitude, $\Delta D$. After getting $D(t)$ and before doing anything else, we also randomly move each centroid to simulate imperfect centering. Here we use a bivariate Gaussian with standard deviations of 1\% the centroid cluster's size in $x$ and $y$, and a random correlation between $[-0.5,0.5]$. These shifts are a little smaller than in \citet{ingalls2014using} and do not strongly affect our results.

Real \emph{Spitzer} data show a variety of intra-pixel sensitivity variations in the different IRAC channels \citep[e.g.][]{stevenson2012transit,triaud2015wasp}. For the example in Figure \ref{fig:samp_cents_pixsens}, the detector sensitivity varies about an order of magnitude more than the eclipse depth we model (below). We will test a range of other scenarios in Sections \ref{sec:map_compare} and \ref{sec:more_synth_fits}.

\subsection{Astrophysical Model}
\label{sec:A_model}
The astrophysical signals we are interested in are planetary eclipses, and we use hot Jupiters as the model because these are the planets that BLISS is often used for. We assume our planets are on circular orbits and only consider thermal emission. Hot Jupiters exhibit thermal phase variations \citep[e.g.][]{knutson2007map,crossfield2012spitzer,wong2015W14b}, which we model as a sinusoid, $\Phi(t)$:
\begin{equation}
	\label{eq:phase_func}
	\Phi(t) = 1 - \alpha \cos\left(\frac{2\pi}{P_{\mathrm{orb}}} t + \phi_{o}\right),
\end{equation}
where $\alpha$ is the half-amplitude, $P_{\mathrm{orb}}$ is the orbital period, $t$ is the time from the start of the observation, and $\phi_{o}$ is the phase offset. The constant keeps $\Phi(t)$ close to unity, and we fix $t_{\mathrm{max}} = 6$ hrs because real observations are about that long.

Then we inject the eclipse to get the full astrophysical model, $A(t)$:
\begin{equation}
	\label{eq:A_signal}
	A(t)=\begin{cases}
		\overline{\lbrace\Phi(t) - \delta_{e}\rbrace}_{\textrm{eclipse}}, & |t - t_{e}| \leq t_{w}.\\
		\Phi(t), & \mathrm{otherwise},
	\end{cases}
\end{equation}
where $\delta_{e}$ is the eclipse depth, $t_{e}$ is the time at the center of eclipse, and $t_{w}$ is the time from $t_{e}$ to ingress or egress. We choose $t_{w} = 1$ hr because real eclipses of hot Jupiters usually last a couple hours. The bar in Equation \ref{eq:A_signal} means we take the average of all data during the eclipse, so ingress and egress are instantaneous and the bottom of the eclipse is flat. The details about how we choose the other parameters for $A(t)$ are given in Appendix \ref{sec:A_param_explain}.

Finally, we combine Equations \ref{eq:D_signal} and \ref{eq:A_signal} to create our model of \emph{Spitzer} light curves:
\begin{equation}
	\label{eq:flux_model}
	F(t) = A(t) D(t) + N(t;\sigma),
\end{equation}
where $F(t)$ is the flux, $D(t)$ is the detector signal in Section \ref{sec:D_model}, and $N(t;\sigma)$ is photon (Gaussian) noise with uncertainty $\sigma$. We characterize our light curves using the normalized detector amplitude, $\Delta D/\delta_{e} \equiv \Delta D_{e}$, and the significance of the eclipse, $\mathbb{S}_{e}$, defined as:
\begin{equation}
	\label{eq:signif_ecl}
	\mathbb{S}_{e} \equiv \frac{\delta_{e} \sqrt{N_{e}}}{\sigma},
\end{equation}
where $N_{e}$ is the number of data during the eclipse.

\begin{figure}[h!]
	\centering
	\includegraphics[width=0.75\linewidth]{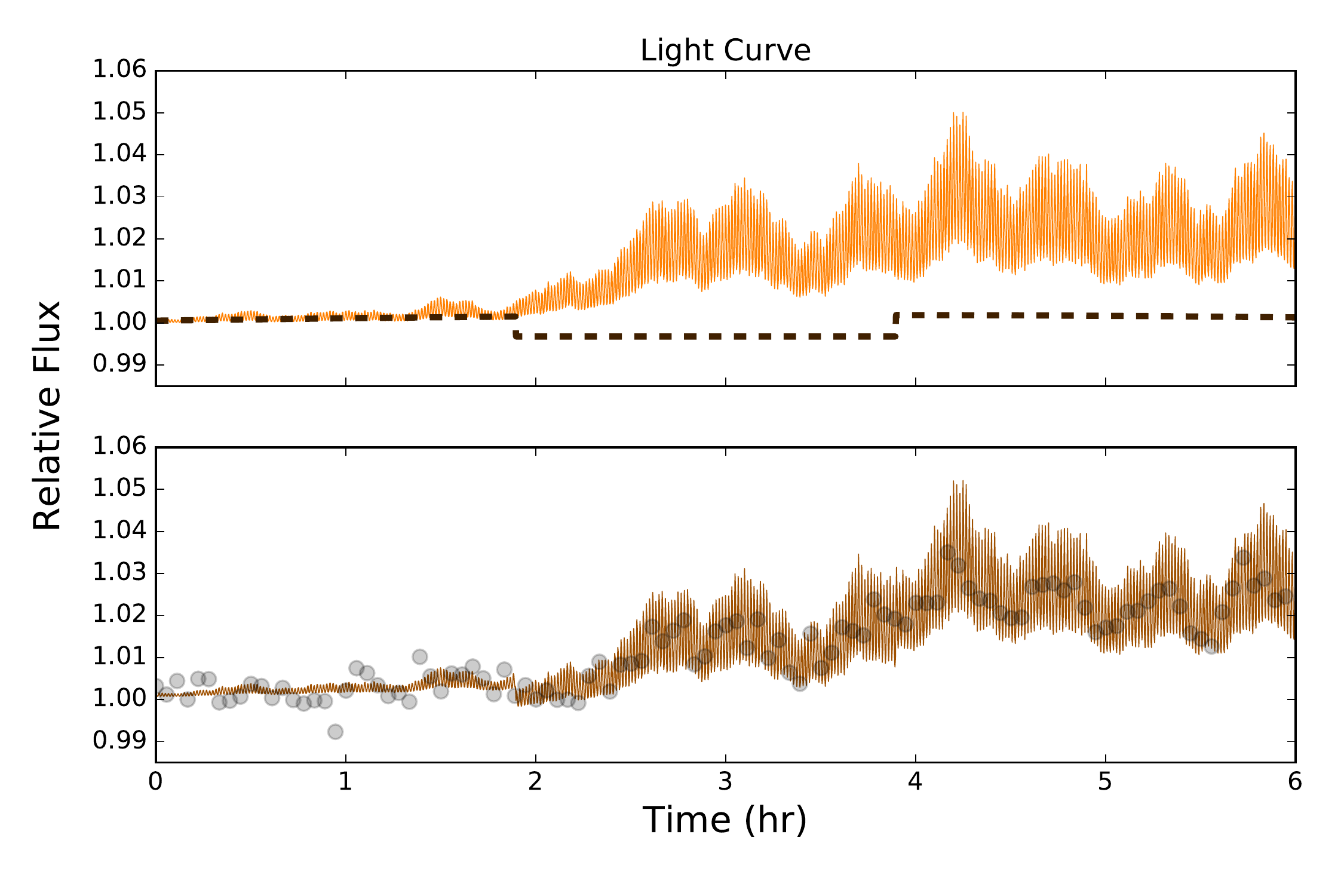}
	\caption[A mock light curve and its two components.]{\emph{Upper:} Examples of the two components in light curves that we model. The dark dashed curve is an astrophysical signal made with Equation \ref{eq:A_signal}. The orange curve shows a detector signal made with Equation \ref{eq:D_signal}, using the centroids and sensitivity map in Figure \ref{fig:samp_cents_pixsens}. The amplitude of this detector signal is $10\times$ larger than the eclipse depth---we test other cases in Sections \ref{sec:map_compare} and \ref{sec:more_synth_fits}. \emph{Lower:} A synthetic light curve made with Equation \ref{eq:flux_model} and the above signals. The brown curve shows the flux with no photon noise and the gray circles are data, binned in groups of 20 for clarity. The eclipse is a $10\sigma$ detection for these data.}
	\label{fig:samp_lightcurve}
\end{figure}

An example light curve is shown in Figure \ref{fig:samp_lightcurve}, made with the centroids and sensitivity map in Figure \ref{fig:samp_cents_pixsens}. The upper panel shows the astrophysical and detector signals as a dark dashed curve and an orange curve, respectively. These parts are combined in the lower panel: the brown curve is the flux one would see without photon noise, and the gray circles are data points, binned in groups of 20 for clarity. For this case, the eclipse is detected at $10\sigma$ and $D(t)$ has an amplitude $10\times$ larger than the eclipse depth. This type of detector signal is similar to IRAC data at $3.6~\mu$m \citep[e.g.][]{stevenson2012transit,cubillos2013wasp}---we test a variety of values for $\Delta D_{e}$ in Sections \ref{sec:map_compare} and \ref{sec:more_synth_fits}.

%\begin{sidewaysfigure}
%\end{sidewaysfigure}

\section{Tests of BLISS}
\label{sec:B_tests}

\subsection{BLISS Method}
\label{sec:B_routine}
We give a brief summary of BLISS here---for details, see \citet{stevenson2012transit}. A light curve has two main parts: a detector signal (e.g. due to varying sensitivity on the pixel) and an astrophysical signal (e.g. a planetary eclipse). If one knew the astrophysical part and \emph{divided} it out of the light curve, all that should be left in the residuals is the detector signal and photon noise.

\begin{figure}[h!]
	\centering
	\includegraphics[width=1.0\linewidth]{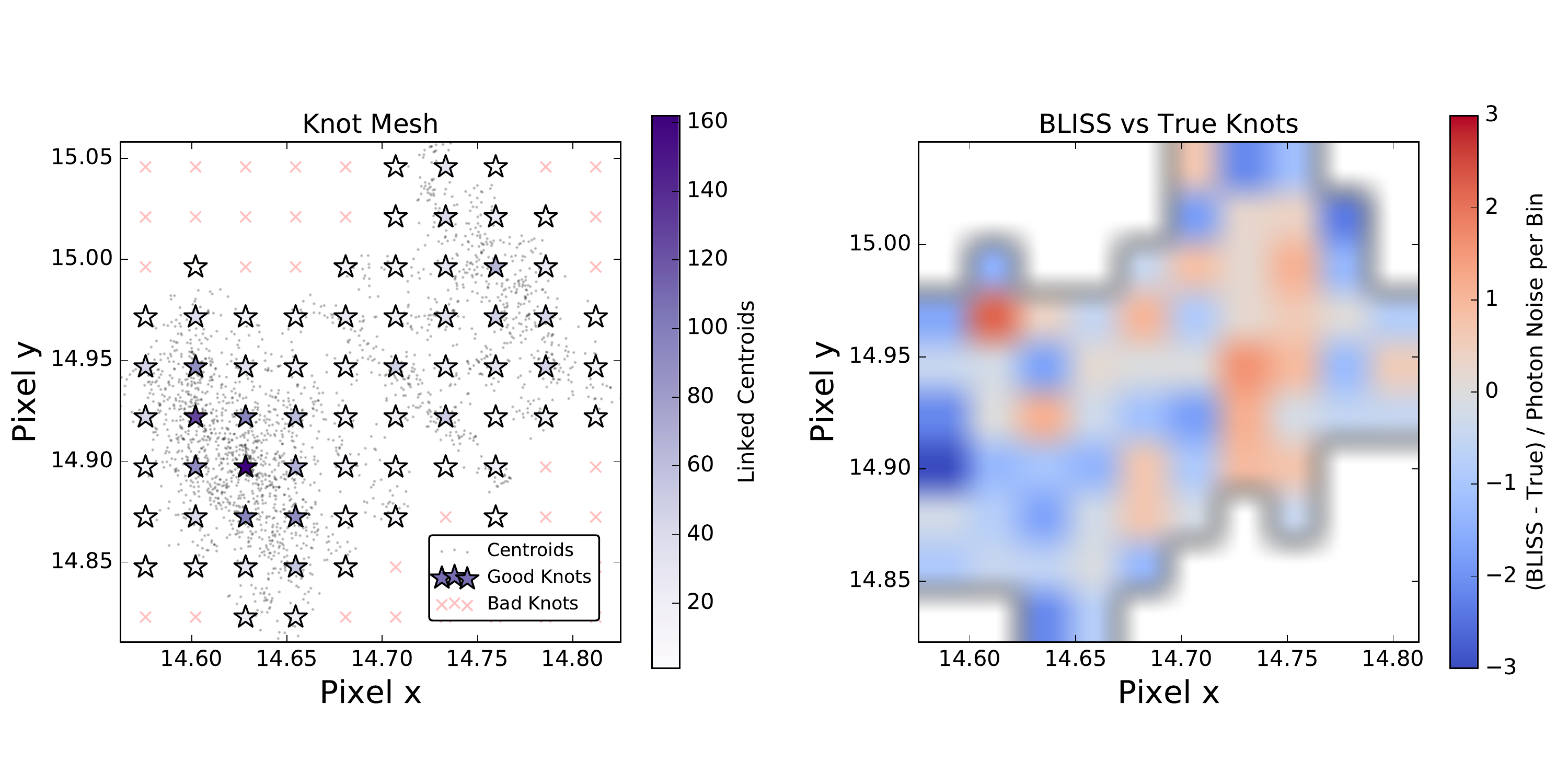}
	\caption[Example BLISS knots and how the BLISS map differs from the true sensitivity.]{\emph{Left:} A $K=10^{2}$ mesh of BLISS knots covering the centroids (gray dots) from Figure \ref{fig:samp_cents_pixsens} (these knots are chosen reasonably, but see Sections \ref{sec:choose_mesh} and \ref{sec:more_synth_fits}). The stars are good knots, or those with at least one centroid nearby, while light red x-marks are bad knots. The color scale shows how many centroids are linked with each knot; darker purple stars are knots where more data is averaged to guess the sensitivity there. BLISS then interpolates the sensitivity at each centroid using the four surrounding knots. \emph{Right:} Discrepancies between the BLISS and true knots, using Equation \ref{eq:knot_discrep}. Darker reds (blues) are where a BLISS knot has a higher (lower) sensitivity than the pixel at that spot; the color scale shows up to $\pm3$. The average discrepancy is about $-0.34$, while the standard deviation is roughly $1.02$---very close to the expected RMS value of unity. These knots also generate a good map (i.e. detector signal) compared to the residuals, with $\chi^{2}/N\approx1.05$. BLISS knots and maps are usually accurate for the data in a light curve, but grow inaccurate when the photon noise is low.}
	\label{fig:samp_mesh_BvsTrue}
\end{figure}

Each residual is paired with a centroid, so one can group the residuals with a mesh of BLISS ``knots," $K$ (left panel of Figure \ref{fig:samp_mesh_BvsTrue}), take the average of each group, and set the values of the knots to these averages. This estimates what the sensitivity looks like on the pixel around the centroids, and each purple star in Figure \ref{fig:samp_mesh_BvsTrue} is a good BLISS knot, or one that has at least one centroid nearby. Other studies \citep[e.g.][]{stevenson2012two,blecic2014spitzer} often require good knots to have at least four linked centroids---those with just one nearby centroid will fit noise by definition. But, this should only affect a tiny part of the detector model and so is negligible. We explicitly try making $K=10^{2}$ an ideal mesh size for our example, but this is difficult to do (Sections \ref{sec:choose_mesh} and \ref{sec:more_synth_fits}).

%\begin{sidewaysfigure}
%\end{sidewaysfigure}

To figure out what $D(t)$ is, BLISS interpolates the sensitivity at each centroid by using the four surrounding knots (hence bilinear interpolation). For centroids where any of those four knots are unconstrained by the residuals (light red x-marks in Figure \ref{fig:samp_mesh_BvsTrue}), BLISS does nearest neighbor interpolation (NNI) instead. Usually a few of our centroids are just outside the mesh of BLISS knots, so we extrapolate the sensitivity at those spots when we can. During the course of an MCMC, a new astrophysical signal is made at each step, the new residuals are averaged, and the detector signal is recalculated. Thus, BLISS tries to attribute unfitted variations to the detector.

\subsection{Comparing Knots and Maps}
\label{sec:map_compare}
BLISS has been used many times to handle sensitivity variations in IRAC data \citep[e.g.][]{diamondlowe2014new,triaud2015wasp,stevenson2016search}, and has been shown to be reliable and accurate at estimating the eclipse depths of XO-3b \citep{ingalls2016repeatability}. But, no research has looked at the accuracy of BLISS knots or maps. We first calculate the true sensitivity at each knot's location by evaluating Equation \ref{eq:sens_vars} there---these are the values that BLISS tries to estimate.

To guess what the best-fit BLISS knots would be, we next take $F(t)/A(t)$ in Equation \ref{eq:flux_model} and use those residuals in the BLISS routine (this estimate is good; Section \ref{sec:synthetic_fits}). Then we compare the BLISS and true knot values to each other:
\begin{equation}
	\label{eq:knot_discrep}
	\delta k_{i} = \frac{(k_{B} - k_{T})_{i}}{\sigma / \sqrt{N_{i}}},
\end{equation}
where $\delta k_{i}$ is the discrepancy of knot $i$, $k_{B}$ is the value of a BLISS knot, $k_{T}$ is the true sensitivity at the same knot, and $N_{i}$ is the number of centroids linked to that knot. The denominator in Equation \ref{eq:knot_discrep} is the photon noise per bin (assuming Poisson statistics), which implicitly weights the discrepancies by the data per \mbox{knot (i.e. star} color in Figure \ref{fig:samp_mesh_BvsTrue}). Again, $\delta k$ measures how well BLISS estimates the sensitivity at the knots---we test the full map, or the interpolated detector signal, further below. We show values of $\delta k$ for our example knots in the right panel of Figure \ref{fig:samp_mesh_BvsTrue}. Although the astrophysical model is known perfectly here, the larger discrepancies can occur in the interior of the mesh where there are more data per knot.

The standard deviation of Equation \ref{eq:knot_discrep} for all the knots tells us how reliable these estimated sensitivities are---average discrepancy matters less because \emph{Spitzer} is poor for absolute photometry of planetary eclipses \citep[e.g.][]{reach2005absolute}. Similar to z-scores in Section \ref{sec:toy_fits}, we expect an RMS value close to unity if the knots are accurate. For example, the standard deviation on $\delta k$ in Figure \ref{fig:samp_mesh_BvsTrue} is about $1.02$ (average is around $-0.34$), so these BLISS knots are indeed a good match to this pixel's true sensitivity. We test this for other light curves by varying four parameters: the number of data points ($N$), the total amount of BLISS knots ($K$), the eclipse significance, and the normalized detector amplitude.

We start by making an $11\times11\times50\times50$ logarithmically-spaced grid of $N\in[10^{2},\sim10^{5}]$, $K\in[5^{2},160^{2}]$, $\mathbb{S}_{e} \in [1,100]$, and $\Delta D_{e} \in [0.1,100]$, respectively. We also try a second grid where the dimensions are reversed. Then we make 3 light curves (Equation \ref{eq:flux_model}) at each grid point, get the BLISS knots as described above, and use Equation \ref{eq:knot_discrep} to find the average standard deviation of $\delta k$. In general, we find a trend in RMS values with $\mathbb{S}_{e} \Delta D_{e} = \Delta D/(\sigma/\sqrt{N_{e}})$, which is the detector amplitude relative to the astrophysical precision on eclipse timescales (Section \ref{sec:A_model}). We also find a similar trend with the average data per BLISS knot, $N/K$. However, the number of \emph{good} knots for given data depends on the shape of the centroid cluster, so we focus more on $\mathbb{S}_{e} \Delta D_{e}$.
 
For given amounts of data and knots, when $\mathbb{S}_{e} \Delta D_{e}$ is low the photon noise is much bigger than the detector amplitude, and the standard deviation of $\delta k$ is around unity. In these cases a BLISS knot is generally as accurate as the noise in the residuals. As $\mathbb{S}_{e} \Delta D_{e}$ goes up, the photon noise decreases, and the knots get closer to the true sensitivities while still being noise-limited. We eventually find an ideal regime, covering about an order of magnitude in $\mathbb{S}_{e} \Delta D_{e}$, where BLISS knots have values similar to the pixel's true sensitivity \emph{and} the RMS of $\delta k$ stays around unity. Above this range, however, the photon noise decreases so much that the standard deviation of $\delta k$ grows, even though the knots stay close to their true values. These are bad levels of $\mathbb{S}_{e} \Delta D_{e}$ because BLISS is not estimating the sensitivity at the knots correctly for the expected precision. Since $N/K$ has a similar trend, this means that when $\mathbb{S}_{e} \Delta D_{e}$ is high for given $N$, BLISS will estimate the sensitivity better by using more knots (i.e. smaller bins).

The full maps (i.e. sensitivity at each centroid) are comparable. When we use the BLISS and true knots to interpolate $D(t)$ for a set of centroids, both typically fit the residuals equally well (i.e. similar Chi-square values) when $\mathbb{S}_{e} \Delta D_{e}$ is in or below the ideal range. This happens in Figure \ref{fig:samp_mesh_BvsTrue}, where both $D(t)$ would have $\chi^{2}/N\approx1.05$. Once $\mathbb{S}_{e} \Delta D_{e}$ is $\sim2\times$ the ideal limit for accurate knots or higher, BLISS maps usually do a little better, but both fits start to become poor. The photon noise is low in these cases, and neither map models the detector signal to within the precision of the data. On the other hand, having $N/K\sim10$ or less means that $\chi^{2}/N < 1$ and the BLISS maps fit progressively more noise. Still, in most cases modeling $D(t)$ with BLISS is statistically as good as interpolating from the true sensitivity at the knots.

For example, with $N\approx2.5\times10^{4}$ and $K\approx30^{2}$, we get ideal BLISS knots when $\mathbb{S}_{e} \Delta D_{e} \in [10,250]$ and good detector signals when $\mathbb{S}_{e} \Delta D_{e} < 500$, both roughly. We can use these values to guess how accurate the BLISS knots and maps are for the studies in Table \ref{tab:BLISS_planets}, which often use similar $N$ and $K$ for eclipse observations. We estimate the detector amplitudes from uncorrected light curves or the sensitivity maps if shown, and the eclipse significances from binned light curves that have uncertainty bars. In general, we find that most studies \citep[e.g.][]{cubillos2013wasp,stevenson2014deciphering} have $\mathbb{S}_{e} \Delta D_{e}$ values within our ideal range---these BLISS knots and maps should be accurate. Two 3.6 $\mu$m cases to note are \citet{blecic2013thermal}, where we estimate $\mathbb{S}_{e} \Delta D_{e} \in [250,450]$ for WASP-14b, and \citet{stevenson2012transit}, with $\mathbb{S}_{e} \Delta D_{e} \in [360,600]$ for HD 149026b. In these studies, the sensitivity at the BLISS knots is likely starting to go bad ($1.0\leq\textrm{RMS}[\delta k]\leq1.5$), but the detector signals should still be modeled well. Naturally, higher values of $\mathbb{S}_{e} \Delta D_{e}$ would be worse.

\citet{stevenson2012transit} states that, when possible, one should choose a bin size for BLISS (i.e. number of knots for given data) which does not depend on the eclipse depth and gives less scatter in the best-fit residuals than NNI. For a given light curve, it seems that one could also use $\mathbb{S}_{e} \Delta D_{e}$ (or $N/K$) to select an ideal number of knots for their BLISS routine. However, any of these guidelines are likely problematic (Sections \ref{sec:choose_mesh} and \ref{sec:more_synth_fits}).

We will test different sizes for the knot mesh when fitting some of our synthetic data with BLISS (Section \ref{sec:more_synth_fits}). Though we will make practical choices for $N$ and $K$ to run MCMC on our light curves, these data mimic published studies and our results should apply to real \emph{Spitzer} observations.

\subsection{MCMC Fits to Synthetic Eclipses}
\label{sec:synthetic_fits}
We want to fit our light curves using MCMC and BLISS, but cannot use lots of BLISS knots because \texttt{emcee} would run very slowly with that many parameters. Indeed, this is why \citet{stevenson2012transit} introduced this residual optimization scheme in the first place. Instead, we start with $N=2160$ and test for the number of BLISS knots to use, suggested by \citet{stevenson2012transit} above.

\subsubsection{Selecting the BLISS Mesh}
\label{sec:choose_mesh}
As stated in Section \ref{sec:A_model}, the data in our main example (a $10\sigma$ eclipse with $\Delta D=10\delta_{e}$; Figure \ref{fig:samp_lightcurve}) is modeled on IRAC at $3.6~\mu$m. We therefore use Table 2 of \citet{stevenson2012transit}, also for $3.6~\mu$m data, as a guide (T2 for short). When we fit light curves like in Figure \ref{fig:samp_lightcurve} with BLISS, the eclipse depths are usually very consistent at $1\sigma$ for $K\in[7^{2},20^{2}]$. Our centroid clusters are $\sim0.2$ pixels wide in $x$ and $y$, so these knots are spaced about $[0.03,0.01]$ pixels apart. This matches T2 well and shows our eclipse depth does not depend on bin size.

Having the best-fit residuals be less scattered for BLISS than NNI is harder to do. One can only ensure this by explicitly fitting a light curve with both methods, not feasible for our study. Instead we approximate these fits by using $F(t)/A(t)$ from Equation \ref{eq:flux_model} in both routines, as in Section \ref{sec:map_compare}. Then from T2, we compare the \emph{ratio} of standard deviations in the best-fit residuals for BLISS and NNI, $\mathbb{R}^{B}_{N}$. We estimate that eclipse depths fit by BLISS in T2 become inconsistent when $\mathbb{R}^{B}_{N}$ drops below $\approx0.987$, at a bin size of $\sim0.06$ pixels. So that our knots are spaced closer than this, we keep $K=10^{2}$ as the starting mesh (i.e. BLISS bin size of $\sim0.02$ pixels) and only use light curves (Section \ref{sec:more_synth_fits}) where we estimate $\mathbb{R}^{B}_{N}\in[0.99,1.0)$. This is our conservative attempt to have BLISS work better than NNI---true for our main light curve in Figure \ref{fig:samp_lightcurve}.

However, our choice is probably arbitrary. The value of $\mathbb{R}^{B}_{N}$ seems to depend on many aspects of a light curve, especially the detailed shape of the detector signal. Worse, when we draw new Gaussian noise in Equation \ref{eq:flux_model} while keeping $A(t)$ and $D(t)$ fixed, $\mathbb{R}^{B}_{N}$ can be above or below unity, sometimes with equal chance. That means different photon noise \emph{with the same uncertainty} can make BLISS look good or unnecessary for given data and $K$. Thus, picking the BLISS bin size, according to \citet{stevenson2012transit}, can need fine-tuning.

Also, once NNI outperforms BLISS, \citet{stevenson2012transit} states that this bin size indicates the centering precision for a particular data set. But, our centroids are typically precise at about 5--15\% of the bin size when $\mathbb{R}^{B}_{N}$ goes above unity. Even using the \emph{perfect} locations of all centroids, NNI can still easily do better than BLISS---centering precision is not related to how BLISS performs. Instead, the bin size where NNI starts giving less scattered residuals than BLISS could be related to the length scale of the sensitivity variations. We hypothesize that both of the above issues can happen when fitting real observations.

Nonetheless, there are other benefits to using $K=10^{2}$. Our average data per good BLISS knot is typically within $[25,40]$. These ratios are smaller than Figure 6 of \citet{stevenson2012transit} suggests, but may be similar to other BLISS studies \citep[e.g. Figure 5 of][]{blecic2013thermal}. Also, $N/K=21.6$ and so our BLISS maps will not fit much noise (Section \ref{sec:map_compare}). More importantly, the estimated sensitivity at our knots should be accurate for light curves where the product of the eclipse significance and the normalized detector amplitude is less than $\sim300$. The same is true for the detector signals when this product is less than $\sim600$. As described in Section \ref{sec:map_compare}, these values of $\mathbb{S}_{e} \Delta D_{e}$ are good approximations for those in published papers. In other words, we want our fits to represent a variety of real data while remaining computationally feasible.

\subsubsection{Models and Main Light Curve}
\label{sec:main_LC_fits}
We use three methods to handle the pixel's sensitivity. Since the true sensitivity is generated with a polynomial model, we try polynomial mapping, or $P$-type. Here though, we choose $n=2$ instead of the real $n=7$ to mimic our inexact understanding of the intrinsic detector sensitivity (we test the impact of this choice below). We also use BLISS as described by \citet{stevenson2012transit}, or $B$-type. We further want to fit the knots directly, so we modify BLISS and make each knot a jump parameter inside the MCMC, or $J$-type. Everything else about BLISS is the same in the $B$- and $J$-type methods.

\begin{figure}[h!]
	\centering
	\includegraphics[width=0.75\linewidth]{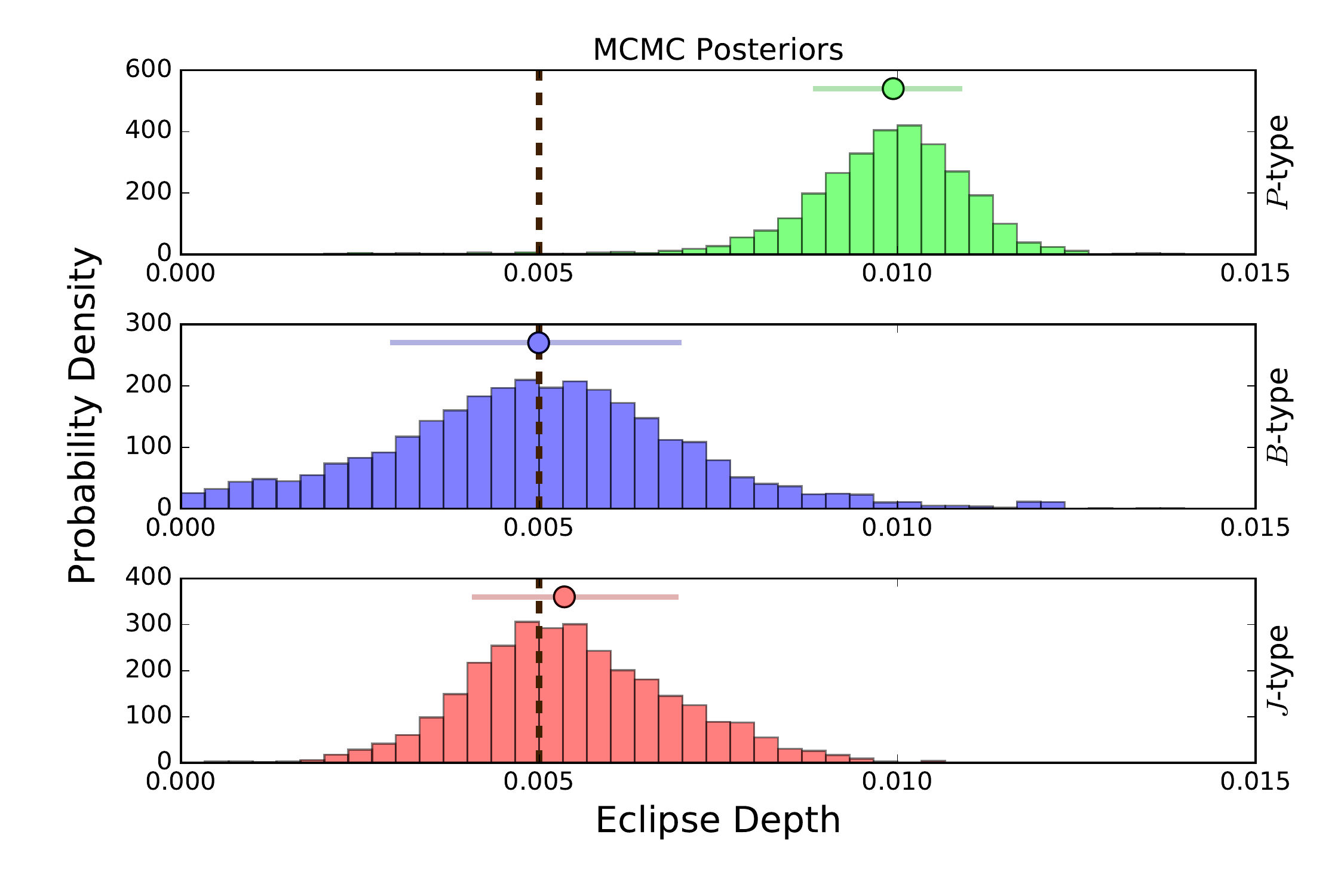}
	\caption[Eclipse depth posteriors for the light curve in Figure \ref{fig:samp_lightcurve}.]{Posterior densities for the eclipse depth in Figure \ref{fig:samp_lightcurve}, fit using MCMC and three models of the sensitivity variations: polynomial or $P$-type at the top (green), BLISS or $B$-type in the middle (blue), and Jump-BLISS or $J$-type at the bottom (red). The dark dashed lines in each panel show the true eclipse depth, $\delta_{e} = 5.0\times10^{-3}$. The circles are medians of each posterior and the bars show the $1\sigma$ intervals. The scales for the eclipse depth are the same, though $\sim0.5\%$ of the $B$-type posterior at larger depths is not shown. Our $\chi^{2}$ values are $2102.9$ for $J$-type, $2173.1$ for $P$-type, and $2182.1$ for $B$-type, so each model has $\chi^{2}/N\approx1$. The polynomial model is the most precise method but both versions of BLISS are more accurate.}
	\label{fig:samp_mainposts}
\end{figure}

We use \texttt{emcee} as in Section \ref{sec:toy_fits}, and for each method ($P$-, $B$-, and $J$-type) we choose the number of walkers to be $3\times$ the number of $J$-type parameters. The priors on all parameters are uniform, and we again start the walkers in a small ball near the true inputs. We run each chain until all parameters stabilize for at least $25\times$ the largest autocorrelation estimate, $\tau_{\mathrm{max}}$, then drop the burn-in and thin the chains by $\tau_{\mathrm{max}}$. Typically, this takes $5$--$20\times10^{3}$ steps and \texttt{emcee} calculates $\tau_{\mathrm{max}} \in [80,100]$ steps. For our example light curve from Figure \ref{fig:samp_lightcurve}, we show all three posteriors on the eclipse depth in Figure \ref{fig:samp_mainposts}. Here the real depth, $\delta_{e} = 5.0\times10^{-3}$, is shown with dashed vertical lines. At the top of each panel, we plot the median depth as a circle and the $1\sigma$ intervals with bars (as in Figures \ref{fig:OptMarg_Posts} and \ref{fig:samp_toy_poly_allMCMC}). Remember that here $\mathbb{S}_{e}=10$ and $\Delta D_{e}=10$.

%\begin{sidewaysfigure}
%\end{sidewaysfigure}

All three posteriors are roughly Gaussian in shape. The $J$-type fit is centered near the true eclipse depth and $B$-type is even closer, but the latter has heavier tails and so is less precise. The $P$-type fit, however, peaks at $\sim2\times$ deeper than the true value. Even though this model is the most precise, it has the worst accuracy. We find that $J$-type has the lowest Chi-square, $2102.9$, compared to $2173.1$ for $P$-type and $2182.1$ for $B$-type. Note that all three models have $\chi^{2}/N\approx1$. As $P$-type shows here, having noisy data can shift best-fit parameters away from their true values, despite Chi-square being good.

\begin{figure}[h!]
	\centering
	\includegraphics[width=1.0\linewidth]{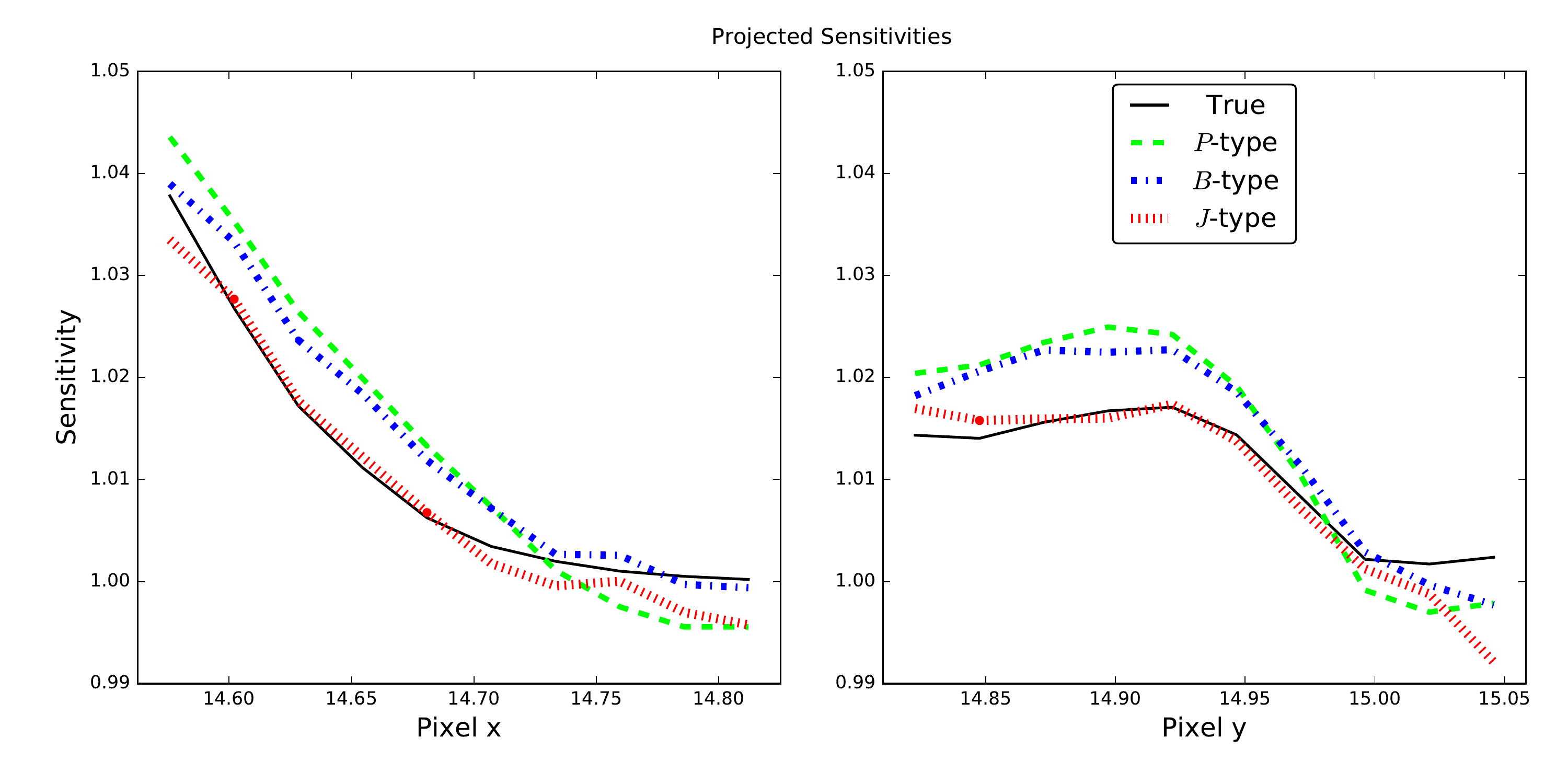}
	\caption[Projected sensitivities for the fits in Figure \ref{fig:samp_mainposts}.]{The projected sensitivities on the $x$- (left) and $y$-axis (right) of the pixel from Figure \ref{fig:samp_cents_pixsens}. The solid black curve is the true result, while the other curves show the best-fit models from Figure \ref{fig:samp_mainposts}: $P$-type in dashed green, $B$-type in dash-dotted blue, and $J$-type in dotted red. All four projections use 10 bins along both axes. Here $J$-type is the model most similar to the true sensitivity variations---BLISS typically matches more featured sensitivities better than polynomials.}
	\label{fig:proj_senses}
\end{figure}

In Figure \ref{fig:proj_senses} we also compare our models to the true sensitivity projected along both axes of the pixel \citep[as in Figure 2 of][]{stevenson2012transit}. Because we use $K=10^{2}$ for BLISS to fit our main light curve, each of these projections is done with 10 bins on an axis. The true sensitivity is shown as a solid black curve, and our models have the same colors as in Figure \ref{fig:samp_mainposts}. As expected from the Chi-square values, $J$-type (dotted red) matches the true variations best, though $P$- (dashed green) and $B$-type (dash-dotted blue) still follow the overall patterns. The astrophysical model can balance out sensitivities here that do not match the true pixel, but this is more helpful for projections that are uniformly high or low along $x$ or $y$. When we try fitting different types of light curves (Section \ref{sec:more_synth_fits}), we find that BLISS is usually better than polynomials at matching more featured kinds of projected sensitivities.

%\begin{sidewaysfigure}
%\end{sidewaysfigure}

As \citet{stevenson2012transit} describes in their Appendix A, comparing BLISS to polynomial models using the Bayesian Information Criterion \citep[BIC; ][]{schwarz1978estimating} is not sound: many parameters do not overlap and each BLISS knot only interacts with a subset of the data. One could try modifying BIC to account for the latter point, but that is beyond the scope of this paper. Instead we will compare models in Section \ref{sec:more_synth_fits} by using accuracy and precision of the fitted eclipse depths.

\subsubsection{Properties of BLISS}
\label{sec:fiduc_BJ}
For the $B$-type model, we find that the best-fit BLISS knots (not shown) are mostly similar to those we estimated in Figure \ref{fig:samp_mesh_BvsTrue}. The $\chi^{2}/N$ for $D(t)$ is higher than our original guess ($\approx1.24$ versus $\approx1.05$), but both signals also look similar. We get the same results when we test other light curves, and that means we can usually estimate the best-fit BLISS knots and map well without running an MCMC. Our findings in Section \ref{sec:map_compare} are therefore robust, and this supports our attempt to choose an ideal BLISS mesh in Section \ref{sec:choose_mesh}.

\begin{figure}[h!]
	\centering
	\includegraphics[width=0.6\linewidth]{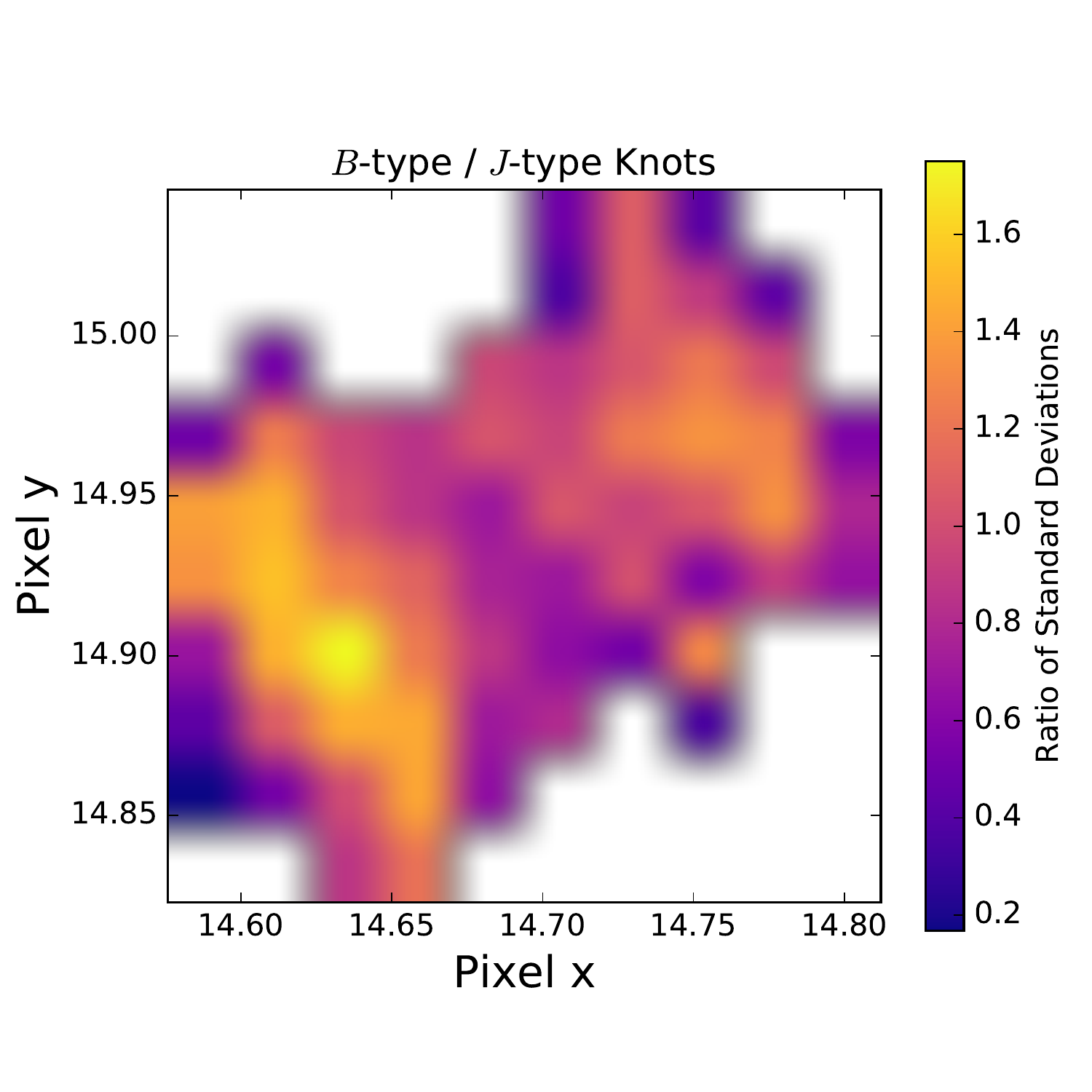}
	\caption[How BLISS knots vary if optimized or fit directly.]{Ratio of standard deviations for knots sampled in the $B$- and $J$-type models. Lighter colors mean those knots varied relatively more in the $B$-type MCMC; the highest ratios are in the interior of the mesh. The value of $B$- and $J$-type knots tend to vary as much during an MCMC, regardless of whether they are jump parameters.}
	\label{fig:samp_knots_STD}
\end{figure}

To test how the BLISS knots vary in the MCMC, we save the knots at every step in the $B$-type model and compare the standard deviation of each $B$- and $J$-type knot in Figure \ref{fig:samp_knots_STD}. Even though the $J$-type knots are free parameters, the $B$-types can vary more (color scale), especially those in the interior of the mesh. This is probably because there are more data per knot here, meaning the central knots have the biggest impact on the detector signal and so vary the most. We see this happen in every light curve we test (i.e. ratio of standard deviations between $[0.3,2.0]$ typically), so in general BLISS knots act like real variables rather than fixed parameters.

We also try slicing through the $J$-type MCMC chain (i.e. posterior; Section \ref{sec:marg_opt}) in the knot parameters. This shows how the fit to the eclipse depth changes when fixing the knot values, which should be a worst-case scenario for BLISS. With $65$ good knots, though, this is tricky. For example, the density of a $\nu$-dimensional Gaussian depends on the $\sigma$-scaled distance $d$ from the mean \citep[Mahalanobis distance; e.g.][]{de2000mahalanobis}, where $d^{2}$ has a $\chi_{\nu}^{2}$ distribution \citep[e.g.][]{tong2012multivariate}. The chance that a point lies within $d=1$ in our case, or $1\sigma$ in \emph{all} knots, can be estimated as $\mathrm{CDF}(1^{2};\chi_{65}^{2})\sim10^{-48}$. Our thinned chains only have $\sim10^{4}$ samples, so it is near-impossible to have any sample close to the best-fit value of every knot. Discrete samples are often \emph{very} spread out in a high-dimensional space.

In practice we take slices much larger than $1\sigma$ through the $J$-type knots to capture close to $10\%$ of the samples. The above example predicts this happens when $d\approx7.13$. When we slice around the maximum likelihood value of the knots, we only need $d\approx2.42$. The median eclipse depth is about $4\%$ higher than in the full chain, and the interval in nearly unchanged. If we slice around the median knot values instead, we only need $d\approx2.04$. The median eclipse depth goes down by $\sim5\%$, but the interval is now about $18\%$ smaller. The low $d$-values we find imply that the $J$-type posterior is not a multivariate Gaussian.

When we test other light curves, the $J$-type slices often look similar. Median eclipse depths are usually within $10\%$ of those in the full chains (i.e. good for $10\sigma$ eclipses or better), and the intervals between $20\%$ narrower to $10\%$ \emph{wider}. The exceptions are when the eclipse significance is low: these fit intervals on the depth are around half the width of those in the full chains. But generally, slicing through the $J$-type posterior---which limits the value of each knot---does not affect the fitted eclipse depth much.

\subsubsection{Varying the Data and BLISS}
\label{sec:more_synth_fits}
So far we have (mostly) considered the fits for a single light curve. We now try fitting different data sets and changing how many BLISS knots we use. For consistency, we fix all eclipse depths to $\delta_{e} = 5.0\times10^{-3}$ and test 5 light curves, or 10 where stated, per case we consider. We randomly generate these synthetic data, but visually inspect them to make sure the detector signal is not mostly flat, which happens about 10--20\% of the time. Note that we only explicitly try to have BLISS work better than NNI (Section \ref{sec:choose_mesh}) in our main type of light curves (circles below; includes Figure \ref{fig:samp_mainposts}) and when we later modify the sensitivity variations for this type. Other cases are experiments on changing some aspect of the data or BLISS.

Because we find above that the $B$- and $J$-type models are similar, we drop $J$-type from here on to speed up our fits. We repeat the $P$- and $B$-type MCMCs as described in Section \ref{sec:main_LC_fits}, and since there are several changes to consider, we split these sets of light curves into groups. We find all $\chi^{2}/N\sim1$ and either model can have the lowest value unless stated otherwise. Bear in mind that the following figures only show about a third of our MCMC fits---we have tried other (sometimes uninteresting) parts of the parameter space.

\begin{figure}[h!]
	\centering
	\includegraphics[width=1.0\linewidth]{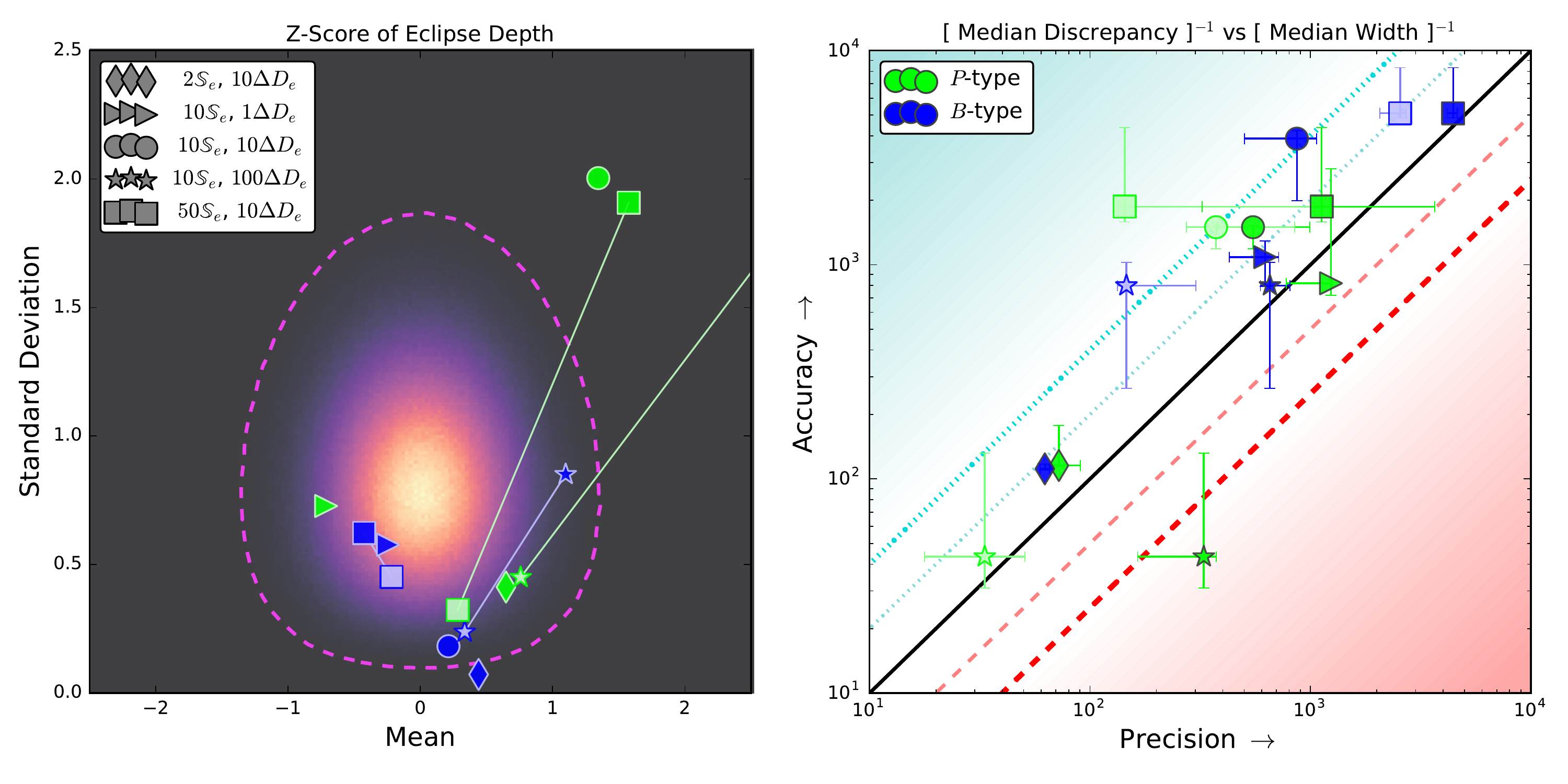}
	\caption[Accuracy/precision of eclipse depths: varied eclipse significances and detector amplitudes.]{\emph{Left:} Mean values and standard deviations of the z-scores obtained by fitting the eclipse depth in different types of light curves; here we vary the eclipse significance and normalized detector amplitude. Each marker uses 5 samples---the fits from Figure \ref{fig:samp_mainposts} are part of the circles---with the $P$-type data shown in green and the $B$-type in blue. Lighter markers show, when significant, how the z-scores change by accounting for time-correlated residuals via the $\beta$ method (Section \ref{sec:more_synth_fits}), where the connecting lines are for clarity. The original green star is outside the plot at roughly $(7.0,4.7)$. The background shows the expected scatter ($10^{7}$ Monte Carlo) for sets of 5 reliable z-scores, where lighter colors are more probable and the dashed magenta line is the 99\% ellipse. \emph{Right:} The reciprocal of both the median discrepancy (i.e. accuracy; z-score numerators) and median width of the fit interval (i.e. precision; z-score denominators) for these eclipse depths. Both axes are logarithmic. The bars show the interquartile ranges in accuracy and precision. Note that inflating uncertainties on the eclipse depth via the $\beta$ method (lighter markers) only affects the precision of the fits. The solid black line shows ideal cases where accuracy and precision are equal (i.e. maximum predictive power). Parallel, the dotted cyan and dashed red lines show where the ratio is $2\times$ and $4\times$ too conservative and too precise, respectively---the cyan and red background shows how these trends continue. BLISS gives better fits in our main case (circles) and when the eclipse significance is higher (squares), but the polynomial model can also do well (e.g. triangles).}
	\label{fig:zscores_accprec_SeDe}
\end{figure}

We first try varying both the eclipse significance and normalized detector amplitude, and show the mean and standard deviation of the z-scores (Equation \ref{eq:z_score}) for the eclipse depth in the left panel of Figure \ref{fig:zscores_accprec_SeDe}. The $P$-type models are colored green and the $B$-types are blue. There are also two kinds of z-scores: the darker markers are the fits, while the lighter markers use more conservative intervals we get by testing for time correlations in the best-fit residuals \citep[$\beta$ plots; e.g.][]{pont2006effect,cowan2012thermalW}. These pairs are clarified with connecting lines and the lighter markers are only shown when they do not overlap the darker version.

%\begin{sidewaysfigure}
%\end{sidewaysfigure}

In Section \ref{sec:toy_fits} we described that parameter fits are reliable if, after many samples, the z-scores on the fits have an average of about zero and a standard deviation around unity. Each marker here only uses 5 samples, so the background shows the scatter we get when drawing, via Monte Carlo, $10^{7}$ sets of 5 samples from a standard normal distribution. Lighter areas are more probable and the dashed magenta ellipse contains 99\% of the Monte Carlo sets. If a marker is outside this region, it likely means the eclipse depths in that case are being fit unreliably. BLISS has some trouble when the eclipse is noisier and the detector signal is larger (blue diamond). The polynomial model has suspect fits when $\mathbb{S}_{e} \Delta D_{e}\geq100$, with the darker green star outside the plot at about $(7.0,4.7)$. Including $\beta$ factors makes the z-scores reasonable for the higher two cases (lighter green square and star).

Z-scores combine the accuracy (i.e. discrepancy from a true value; numerator) and precision (i.e. width of an interval; denominator) of each \emph{individual} fit. By separating these pieces, we can also compare the overall accuracy and precision for \emph{types} of fits. The right panel of Figure \ref{fig:zscores_accprec_SeDe} plots the reciprocal of both the median discrepancy and median interval width---accuracy goes up logarithmically towards the top and precision towards the right. The uncertainty bars show the interquartile ranges when these are larger than the size of the markers. Because we test 5 samples in each case, this means the uncertainty bars ignore the single highest and lowest accuracy and precision we find. Ideally markers will be on or near the solid black line, where accuracy equals precision and the fits have maximum predictive power. If both models are on this line, the one closer to the upper right corner is preferred.

Towards the upper left the fitted eclipse depths are too conservative. The dotted cyan lines show where the accuracy is $2\times$ and $4\times$ larger than the precision (e.g. green square and blue circle). Worse, in the other direction the fits are too confident, the dashed red lines showing where accuracy is $2\times$ and $4\times$ \emph{smaller} than precision. The green star is outside the latter line, but similar to the left panel, accounting for time-correlated residuals (i.e. inflating the uncertainties on the eclipse depth) moves this marker close to the ideal ratio. The blue square and star have reasonable z-scores but are moved off the maximum predictive line by $\beta$ factors---we will return to this point when testing other sensitivity variations later on. These $\beta$ factors can only decrease the precision of the fits; they cannot affect the accuracy.

Relative to our main example (circles), increasing the eclipse significance (squares) helps BLISS more than the polynomial model. In fact, $B$-type is preferred in both these cases due to, respectively, more reliable z-scores or better accuracy and precision. We find $P$-type is the preferred model for the lowest value of $\mathbb{S}_{e} \Delta D_{e}$ (diamonds). For the highest value (stars), the polynomial fits have more predictive power, but are less accurate \emph{and} precise than BLISS. Unexpectedly, $P$-type is at least as precise as $B$-type in three of these five cases (diamonds, triangles, and circles). This is unusual because BLISS has been shown to perform better than a second-order polynomial on real \emph{Spitzer} data \citep[e.g.][]{stevenson2012transit,blecic2013thermal}. BLISS is the more precise model when $\mathbb{S}_{e} \Delta D_{e}\geq500$, though, especially after including $\beta$ factors. Since these values are at or above the limit of $\mathbb{S}_{e} \Delta D_{e}$ for accurate BLISS maps (Section \ref{sec:choose_mesh}), it is not surprising that $\beta$ factors decrease the precision of these fits. But again, the z-scores for the $B$-type square and star are reasonable to start.

\begin{figure}[h!]
	\centering
	\includegraphics[width=1.0\linewidth]{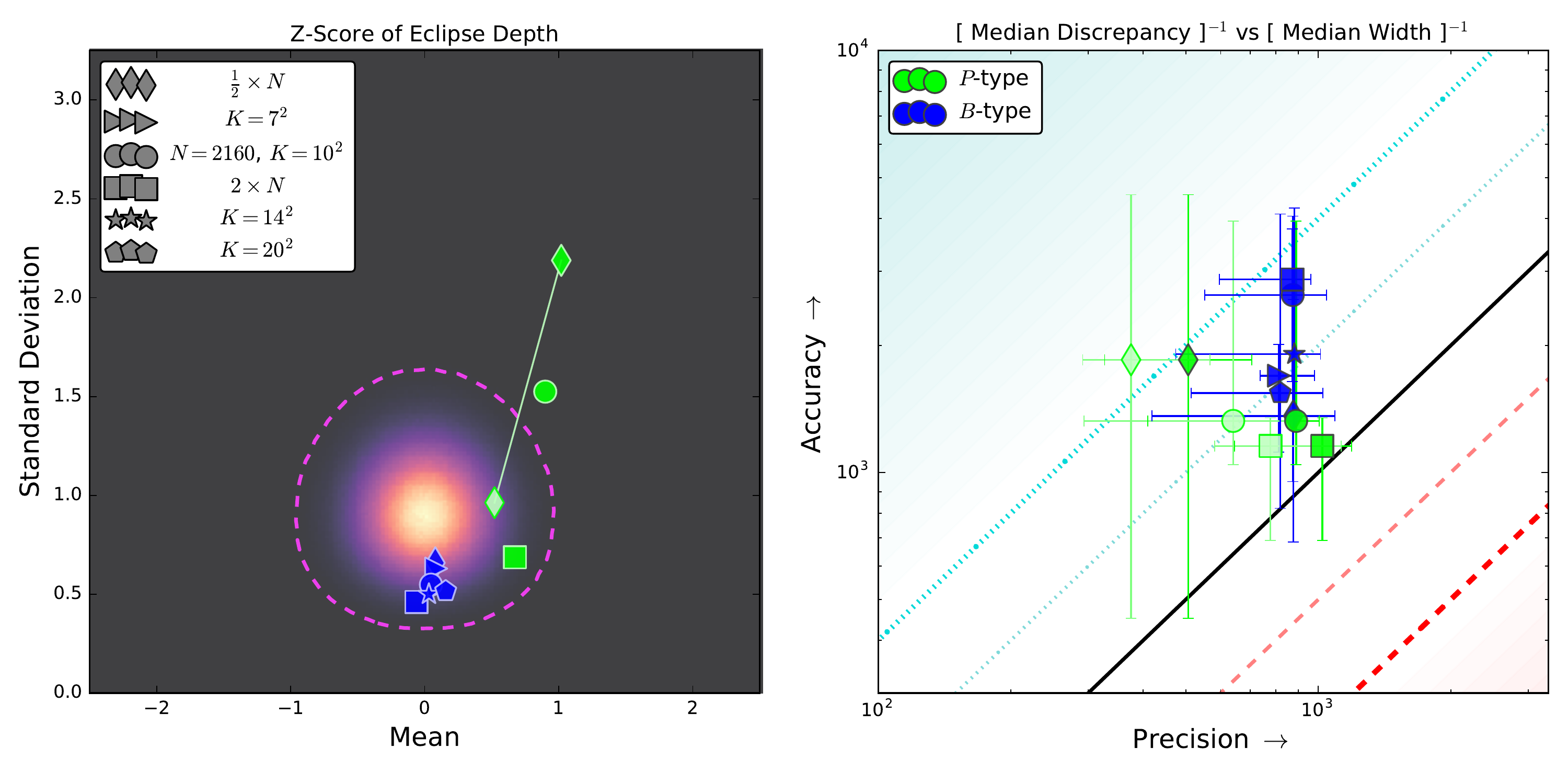}
	\caption[Accuracy/precision of eclipse depths: amount of data and BLISS knots.]{Z-scores, accuracy, and precision of fitted eclipse depths when we vary the amount of data ($N$) or the total number of BLISS knots ($K$). We test 10 samples (instead of 5) in each case here---the background and dashed magenta ellipse in the left panel account for this. The circles include the fits from Figure \ref{fig:zscores_accprec_SeDe}; note that the green circle covers the blue diamond in the right panel. Here the green circle should be compared with the blue triangle, circle, star, and pentagon, which all use the same light curves. When increasing the amount of data (diamonds to circles to squares), BLISS increases in accuracy and the polynomial model moves towards the maximum predictive line (i.e. solid black), though these trends are weak. We confirm that the accuracy and precision of BLISS are consistent when using different numbers of knots, as expected from \citet{stevenson2012transit} and Section \ref{sec:choose_mesh}.}
	\label{fig:zscores_accprec_NK}
\end{figure}

We show similar z-score, accuracy, and precision data as Figure \ref{fig:zscores_accprec_SeDe} for all of the remaining figures. In Figure \ref{fig:zscores_accprec_NK} we test how the fits change with the number of data or BLISS knots, while keeping $\mathbb{S}_{e}=10$ and $\Delta D_{e}=10$. The blue triangle, circle, star, and pentagon use the same 10 light curves, and since changing the BLISS mesh does not affect $P$-type, these fits should be compared to the green circle. The z-scores for both models are acceptable (i.e. markers inside the dashed magenta ellipse) in all new cases after using $\beta$ factors for the green diamond. Note that the blue diamond is behind the green circle in the right panel. In every case the polynomial model and BLISS overlap in precision, given the uncertainty bars.

We see weak trends for both models when varying the amount of data: $P$-type increases in precision yet gets a little less accurate, and $B$-type increases slightly in accuracy. Actually, the polynomial model is preferred when we use more data (squares). Changing the number of BLISS knots affects $B$-type in the right panel, but all four cases mutually overlap in precision and even accuracy. We choose these light curves so that $K=10^{2}$ (blue circle) should be optimal for BLISS (Section \ref{sec:choose_mesh}). However, according to \citet{stevenson2012transit}, the eclipse depths we fit should not depend on the number of knots, unless we make $K$ much smaller. Figure \ref{fig:zscores_accprec_NK} confirms this.

When we also fit some light curves from these cases using NNI (not shown), the best-fit residuals for any $K$ are always less scattered than for BLISS. This does not happen when the eclipse significance is extremely high, or in samples we test from Figure \ref{fig:zscores_accprec_SeDe} with $\mathbb{S}_{e}=50$ and $K=10^{2}$. Yet here, the accuracy and consistency of our $B$-type fits are the same \emph{or better} when compared to NNI. Therefore, our method in Section \ref{sec:choose_mesh} to properly select $K$ for BLISS may have issues (e.g. $\mathbb{R}^{B}_{N}$ must be lower), or the criteria in \citet{stevenson2012transit} may not work in general. Both ideas could be true.

%\begin{sidewaysfigure}
%\end{sidewaysfigure}

\begin{figure}[h!]
	\centering
	\includegraphics[width=1.0\linewidth]{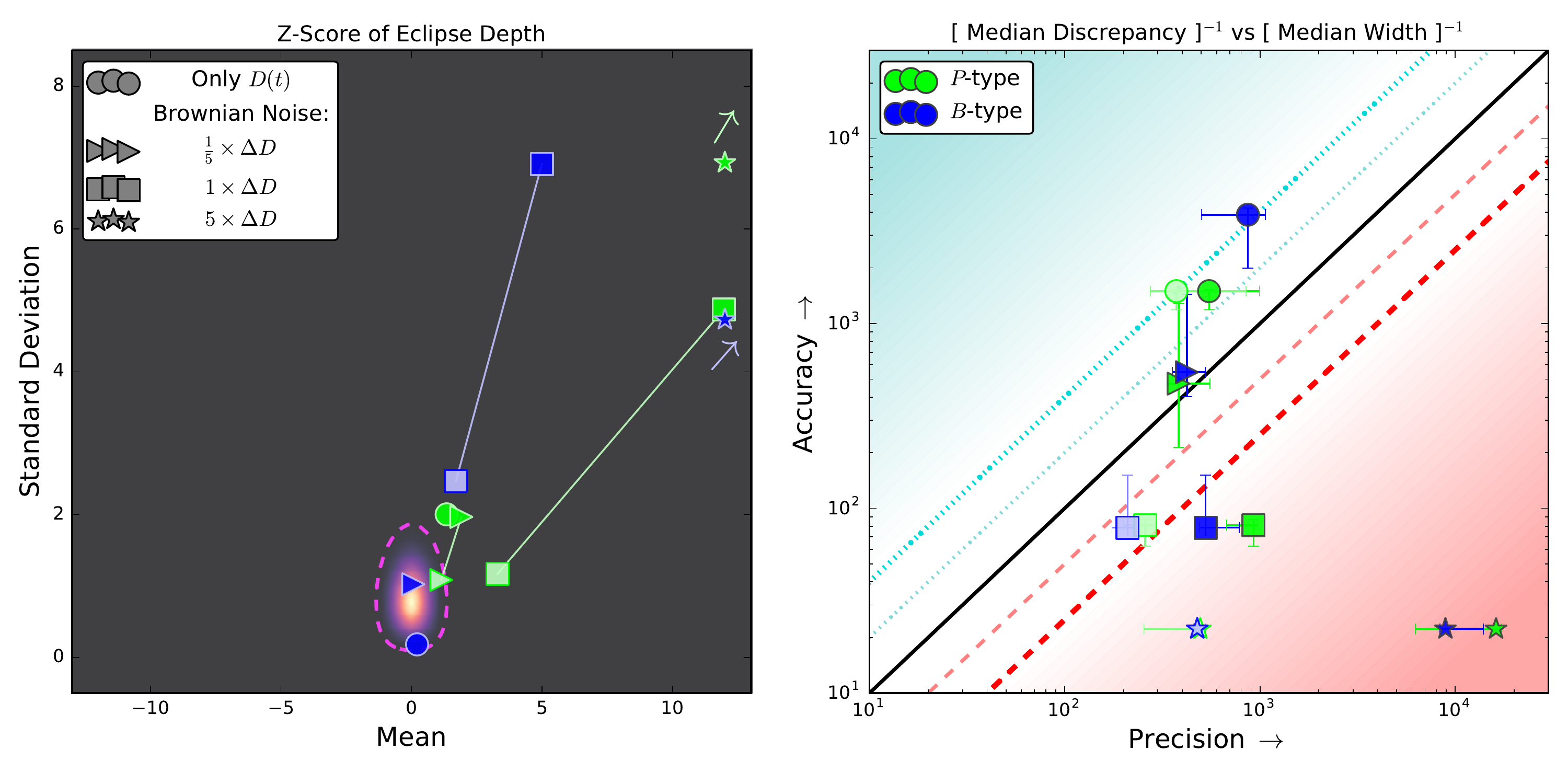}
	\caption[Accuracy/precision of eclipse depths: varied red noise.]{Z-scores, accuracy, and precision of fitted eclipse depths when we multiply different levels of red (i.e. Brownian) noise into the light curve, in terms of the detector amplitude. The circles are taken from Figure \ref{fig:zscores_accprec_SeDe}. All cases use the same 5 light curves and red noises---we only change the amplitudes of the latter relative to their means. Both models are poor in the case with highest noise (stars), where the arrows show that the z-scores are far outside the plot. In the moderate case (squares), the $B$-type fits approach the maximum predictive line with $\beta$ factors included (right), even though these z-scores are outside the 99\% ellipse (left). This means a model that fits eclipse depths well on average may still be unreliable. In all cases here, the polynomial model is as precise as BLISS. Interestingly, having a low amount of red noise in a light curve (triangles) may actually improve the predictive power of both models, especially BLISS.}
	\label{fig:zscores_accprec_Noise}
\end{figure}

Next we test how having more red noise can affect the fits. In Figure \ref{fig:zscores_accprec_Noise} we multiply an extra \mbox{noise (Brownian)} into the light curve to mimic different kinds of time-correlated features (i.e. other than intra-pixel sensitivity variations). We use the same 5 light curves and red noises in each case, meaning we only change the amplitude (relative to the mean) of the noises, not their structure. As often before, $P$- and $B$-type have similar precisions every time.

The case with highest noise, at $5\times$ the detector amplitude (stars), is clearly bad. The z-scores for $P$- and $B$-type are far outside the plot even with $\beta$ factors included ($\sim10$--$35$ on both axes), and the fits are very over-precise. At $1\times$ the detector amplitude (squares), $\beta$ factors move BLISS close to intersecting the maximum predictive line in the right panel, but not inside the 99\% ellipse in the left panel. We find other cases where this outcome is more pronounced, which is a good lesson: the accuracy and precision of a model are separate scalars. Z-scores are a discrepancy \emph{paired} with a fit interval---those specific pairings matter. That means different sets of z-scores can give the same accuracy and precision. Just because a model does well on average does not mean the individual fits are reliable, and vice versa.

The case with extra noise at $\frac{1}{5}\times\Delta D$ is curious (triangles). The z-scores for $B$-type are reasonable, and those for $P$-type are acceptable when using $\beta$ factors. Moreover, both models have near-ideal accuracy and precision. Thus, adding a low amount of time-correlated noise to the synthetic data actually \emph{improves} the predictive power of both fits---especially for BLISS which has insignificant $\beta$ factors. This bodes well for fitting eclipse depths in real light curves because it suggests that one may not need to perfectly model every source of red noise.

%\begin{sidewaysfigure}
%\end{sidewaysfigure}

\begin{figure}[h!]
	\centering
	\includegraphics[width=1.0\linewidth]{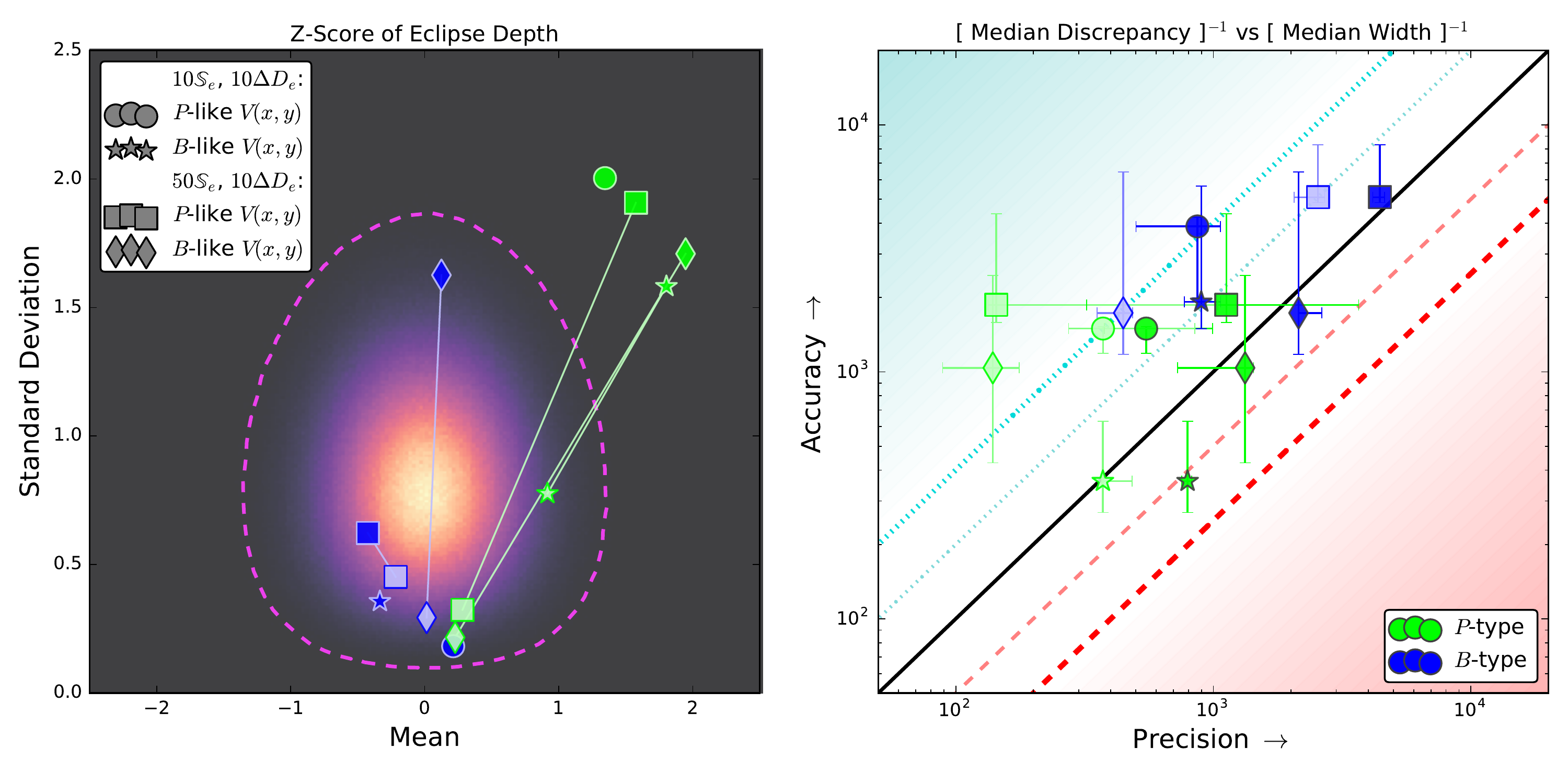}
	\caption[Accuracy/precision of eclipse depths: different sensitivity maps.]{Z-scores, accuracy, and precision of fitted eclipse depths when we use sensitivity variations on the pixel that are either $P$-like (i.e. Equation \ref{eq:sens_vars}) or $B$-like (i.e. defined at the knot locations and interpolated) to create a light curve. The circles and squares are taken from Figure \ref{fig:zscores_accprec_SeDe}. For the case shown by the blue diamond, accounting for time-correlated residuals changes the accuracy and precision from near-ideal (solid black line) to very conservative (outer dotted line and cyan region)---using these $\beta$ factors can be dicey. BLISS is the more precise model for light curves with $B$-like sensitivity variations, but the polynomial fits have more predictive power when these variations are used in our main case (stars).}
	\label{fig:zscores_accprec_Map}
\end{figure}

We further test what happens to the fits when we modify the sensitivity variations on the part of the pixel under the centroids. Note that we already place the centroids at many locations on the pixel (Section \ref{sec:D_model}) to have different terms in Equation \ref{eq:sens_vars} dominate the detector signals (Appendix \ref{sec:D_sens_explain}). In Figure \ref{fig:zscores_accprec_Map} we compare two forms of the pixel's actual sensitivity variations, using two combinations of $\mathbb{S}_{e}$ and $\Delta D_{e}$. The circles and squares (taken from Figure \ref{fig:zscores_accprec_SeDe}) have light curves made with ``$P$-like" variations, or the polynomial $V(x,y)$ in Equation \ref{eq:sens_vars}. The stars and diamonds use ``$B$-like" variations on the pixel instead. For these we define the sensitivity as random Gaussian values at the locations of the BLISS knots. Then we interpolate the sensitivity between these spots using bivariate splines, similar to how the BLISS routine maps $D(t)$ at the centroids.

At first glance BLISS looks like the perfect model for the $B$-like scenario, but it is not. Remember, BLISS estimates the sensitivity at a knot by averaging the residuals (i.e. flux divided by an astrophysical model) at centroids in a bin around that knot. If that bin contains a local peak or valley in the variations, this can throw off the estimate the closer that feature is to the knot. Even when the knot values are accurate, interpolating $D(t)$ well is tricky when the pixel's sensitivity has small-scale structure, especially because the number of knots cannot be increased arbitrarily (Sections \ref{sec:map_compare} and \ref{sec:choose_mesh}). BLISS interpolates linearly between knots adjacent in $x$ or $y$, so we postulate that the routine could only \emph{exactly} match sensitivities that vary like a plane across the pixel. Unfortunately polynomials would also fit exactly in these cases. Nonetheless, BLISS should handle our $B$-like variations better than polynomial models.

When $\mathbb{S}_{e}=10$ and $\Delta D_{e}=10$, both models decrease in accuracy when switching to the $B$-like scenario (circles to stars). We also find that $\beta$ factors are important for the polynomial: it becomes the better predictive model despite BLISS having higher precision. Even stranger is that $P$-type always has a lower $\chi^{2}$ than $B$-type. When $\mathbb{S}_{e}=50$ and $\Delta D_{e}=10$, having $B$-like variations means the median fit for BLISS moves more than for the polynomial model (squares to diamonds). However, with or without $\beta$ factors, BLISS is more precise than $P$-type and is the preferred model.

The blue diamond here is similar to cases from Figure \ref{fig:zscores_accprec_SeDe}. Despite good z-scores, the $\beta$ factors for this model change the fits from ideal to very conservative (i.e. precisions $\sim3$--$10\times$ less than accuracies). This is different from Figure \ref{fig:zscores_accprec_Noise}, where the poor fits show up in the z-scores. In other words, both panels of Figures \ref{fig:zscores_accprec_SeDe}--\ref{fig:zscores_accprec_Map} are important to see how well a model fits a certain type of light curve. Since we find that $\beta$ factors can penalize poor and reasonable fits just as much, using them to tune one's precision is not always wise.

%\begin{sidewaysfigure}
%\end{sidewaysfigure}

To summarize Figures \ref{fig:zscores_accprec_SeDe}--\ref{fig:zscores_accprec_Map}, the eclipse significance and detector amplitude affect the precision and accuracy of a fitted eclipse depth. Changing the amount of data by factors of two weakly affects the accuracy of BLISS and both the accuracy and precision of polynomial models. Using different numbers of BLISS knots gives consistent eclipse depths (as expected), but we find that heuristics for choosing the bin size---here and in \citet{stevenson2012transit}---are questionable. Large amounts of red noise in a light curve are bad, but having low levels can in fact improve both models' fits, particularly BLISS. We find that BLISS fits data with significant eclipses, and (some) light curves made from BLISS-like sensitivity variations, better than the second-order polynomial. Strangely though, the polynomial model is at least as precise as BLISS in many cases we test, and is even preferred in several of them. As for using $\beta$ factors to inflate uncertainties, we get mixed results: these can change dubious fits into near-ideal ones (e.g. green star in Figure \ref{fig:zscores_accprec_SeDe}), but can also make unreliable fits look reasonable (blue square in Figure \ref{fig:zscores_accprec_Noise}) and reliable fits far too conservative (e.g. blue diamond in Figure \ref{fig:zscores_accprec_Map}).

\section{Discussion}
\label{sec:discuss_knot}

\subsection{Kernel Regression}
\label{sec:kern_regress}
We have focused on BLISS because it is easy to adapt the method to a full Jump-type MCMC. Many researchers use BLISS to model intra-pixel sensitivity variations in IRAC data, but there are other non-parametric methods as well. The original approach is Kernel Regression (KR), first used on the GJ 436 system by \citet{ballard2010search}. To measure the transit depth at a known point in a long time-series, the out-of-transit data (i.e. a control) were used to model the detector once at the start of the analysis. This detector model was then used to correct the in-transit data.

Since then, KR has been applied to phase observations, where the signal spans the entire observed baseline and there are no control data \citep{knutson20123,lewis2013orbital}. Researchers have therefore adopted an optimization strategy similar to BLISS: at every MCMC step, the observed flux is divided by the current astrophysical model and KR is applied to the residuals. The KR implemented by \citet{knutson20123} and \citet{lewis2013orbital} also includes the width of the point-spread function, but this does not change the similarity between KR and BLISS. In fact, recent studies have used this width of the point-spread function in tandem with BLISS \citep{lanotte2014global,demory2016variability,demory2016map}.

KR differs superficially from BLISS because it has no obvious detector parameters, making it less clear how to adapt KR to full Jump-type fits. Nonetheless, one can estimate the effective number of parameters as suggested by Footnote 2 of \citet{hansen2014broadband}, typically of order $10^{2}$. Given the conceptual similarities between BLISS and KR, it is possible that our results about the former apply to the latter.

\subsection{Precision of Polynomial Models}
\label{sec:poly_precise}
From the sets of fitted eclipse depths in Figures \ref{fig:zscores_accprec_SeDe}--\ref{fig:zscores_accprec_Map}, it is surprising that a second-degree polynomial is as (or more) precise than BLISS many times. That does not tend to happen with real \emph{Spitzer} data: in both \citet{stevenson2012transit} and \citet{blecic2013thermal}, the BLISS models are more precise than any polynomials the authors test through order $n=6$. In fact, the choice between the models seems so clear that many works in Table \ref{tab:BLISS_planets} do not mention polynomials at all. Remember, BLISS is more precise and often more accurate when we test light curves with significant eclipses (squares in Figure \ref{fig:zscores_accprec_SeDe}) or those made with BLISS-like sensitivity variations (Figure \ref{fig:zscores_accprec_Map}). But in some cases, one is better off modeling the sensitivity with a low-order polynomial---there are several thoughts about why this can happen.

It would be great to fit real IRAC light curves that have $\sim10^{5}$ data with all of these sensitivity models, but as mentioned in Sections \ref{sec:map_compare}--\ref{sec:synthetic_fits}, this is not computationally feasible (more modest IRAC measurements could work, though). Instead we \emph{mimic} these light curves and fits by using realistic parameters for $A(t)$ and $D(t)$, choosing a reasonable BLISS mesh for the synthetic data, and running our MCMC chains until we get many independent samples. But maybe having more data and BLISS knots simply is different, even though the sensitivity at the knot locations and the interpolated maps are mostly accurate in our tests (Section \ref{sec:choose_mesh}). If so, both parameters likely have to increase as we do not see BLISS improve when changing only the data (Figure \ref{fig:zscores_accprec_NK}). This is not because our bin sizes are too large, either. Our BLISS knots are spaced $\sim0.02$ pixels apart in both $x$ and $y$---in other studies this number ranges from smaller \citep[e.g. $3.6~\mu$m data in][]{diamondlowe2014new} to larger \citep[e.g. $5.8~\mu$m data in][]{blecic2013thermal}.

Also, when we set the amplitude of the detector signal in Equation \ref{eq:D_signal}, we do not pick \emph{when} the sensitivity will rise and fall---that would mean explicitly choosing the centroids. Instead, the pointing model \citep{ingalls2016repeatability} and sensitivity map (Appendix \ref{sec:D_sens_explain}) that we use determine how the detector signal looks. If this $D(t)$ is flat with a single large spike or dip at one moment, most data is uncorrupted by the pixel's sensitivity (we avoid these signals for MCMC fits; Section \ref{sec:more_synth_fits}). Moreover, it is mostly chance that the $V(x,y)$ from Equation \ref{eq:sens_vars} is very featured near the centroids. It only happens with particular sets of coefficients, true for any high-order polynomial. This means the sensitivity variations under the centroids can look quadratic \emph{even if} the detector signal looks complex. Either way, our second-order polynomial model could give a good fit. Generating polynomial $V(x,y)$ by selecting their roots, not coefficients, would thus be interesting test cases. When the pixel's sensitivity is BLISS-like none of this should matter, though these variations have other issues (Section \ref{sec:more_synth_fits}).

One possibility comes straight from \citet{stevenson2012transit}: the MCMC would not converge because the model for the exponential ramp had strong, non-linear correlations. To solve this, the authors orthogonalized the ramp parameters, then transformed back to the first model after the MCMC to get the uncertainties. We have no ramping in our synthetic light curves, but similar correlations could happen in our model for $A(t)$. We see evidence of this when we try fitting a DC offset in Equation \ref{eq:phase_func}, which is why we fix the mean of the phase function to unity. There are a couple of problems with this idea, though. Our MCMC chains have little trouble stabilizing, even if \texttt{emcee} takes a long time to get there. Also, we fit the astrophysical signal identically for all three of our sensitivity models. Even if our parameters are not ideal, each fit should be affected the same way by $A(t)$, meaning this is probably not why the polynomial models are more precise.

Above all else, one might say we simply have not tested the ``right" or enough types of light curves. At worst this means we explored some parts of parameter space that differ from real observations. But again, we have chosen a variety of cases based on real IRAC data, fit five or ten examples of each case to have statistics, and even try sensitivity variations more suited to BLISS (Figure \ref{fig:zscores_accprec_Map}). Our results in Section \ref{sec:more_synth_fits} are also only about a third of all our trials; we tested other combinations of parameters. Indeed, BLISS can be more precise than a second-order polynomial, such as by having a significant eclipse, a very large detector amplitude, or BLISS-like variations in sensitivity on the pixel. Yet polynomials are sometimes preferred and the BLISS method is fundamentally the same each time---we find this odd. As mentioned above, testing types of light curves that extend our cases would be a good way to see if any trends here continue in general.

\subsection{Modeling IRAC Noise}
\label{sec:IRAC_noise}
One might ask what Section \ref{sec:synthetic_fits} means for dealing with detector signals in \emph{Spitzer} data. A potential view is to only use non-parametric methods that properly marginalize over the detector behavior, like ICA \citep{waldmann2012cocktail} and Gaussian Processes \citep{gibson2012gaussian}, or rather use viable parametric methods such as PLD \citep[][]{deming2015spitzer,ingalls2016repeatability}. We find a polynomial model is often as precise as BLISS at fitting our synthetic eclipses. For real light curves, one could use high-order polynomials for the sensitivity (e.g. $n\geq7$) and fit every term directly. The number of parameters would be similar to some of our Jump-BLISS models---we find \texttt{emcee} can handle this many jump dimensions.

However, we also have multiple cases (e.g. Figures \ref{fig:zscores_accprec_SeDe} and \ref{fig:zscores_accprec_Map}) that match other studies where BLISS is the more precise choice. Non-parametric models can misfit uncertainties or bias a result (Figure \ref{fig:OptMarg_Posts}), which is disconcerting because one cannot know how accurately BLISS fits real data. We find, though, that BLISS tends to be more accurate than precise (i.e. conservative) at fitting eclipse depths. This result could hold when the routine uses significantly more data and knots (Figure \ref{fig:zscores_accprec_NK}). And while sources of red noise can ruin a fit, a small leftover amount in a light curve \emph{could} be beneficial for the predictive power of BLISS (Figure \ref{fig:zscores_accprec_Noise}).

Yet there are other problems. While $\beta$ factors \citep[e.g.][]{pont2006effect,cowan2012thermalW} are an expedient way to account for time-correlated residuals, these can turn a reasonable uncertainty on an eclipse depth into overly conservative. BLISS does not predict the centering precision of an observation, though we hypothesize the routine may be able to indicate the length scale of the sensitivity variations. Methods to properly size the BLISS mesh, in this paper and \citet{stevenson2012transit}, also may not work as intended (Sections \ref{sec:choose_mesh} and \ref{sec:more_synth_fits}). Furthermore, BLISS is often used with the Photometry for Orbits, Eclipses, and Transits pipeline \citep[POET;][]{stevenson2012transit,cubillos2013wasp}, as in \citet{ingalls2016repeatability}. This proprietary code in part reduces pixelation of the detector by using flux-conserving, interpolated photometry \citep[e.g. Figures 2 and 5 of][]{stevenson2012transit}. But it is unclear if this influences the apparent sensitivity variations on the pixel, and so makes BLISS more necessary to correct for them. Luckily, the large variations at $3.6~\mu$m with IRAC should mean the influence of pixelation, or POET, is more negligible in this channel \citep[e.g. Figure 7 of][]{blecic2013thermal}.

In any case, if a light curve has distinct astrophysical and detector signals, then one could likely use many approaches to model $D(t)$ reasonably. In contrast, a gradual rise and fall in detector sensitivity while observing a planet could be confused with phase variations. As \citet{ingalls2016repeatability} shows, multiple sensitivity models can all fit the same eclipse depth (of XO-3b) well. We expect and see that this sometimes happens with our synthetic light curves.

There probably is no ideal method for handling the sensitivity variations in \emph{Spitzer} IRAC data, and BLISS has both positive and negative qualities. We deem that the good significantly outweighs the bad, though, and suggest that using BLISS as a shortcut can be a practical approach.

\section{Conclusions}
\label{sec:conclude_knot}
We have performed MCMC fits on synthetic eclipse data to test how accurate and precise BLISS mapping is for modeling intra-pixel sensitivity variations in \emph{Spitzer} IRAC light curves. BLISS mapping is a non-parametric method, meaning it uses no jump parameters during the MCMC to model the detector signal. This is an expedient approximation that is not statistically sound in principle. Nonetheless, BLISS mapping has been widely used without rigorous testing on synthetic data.

Optimizing nuisance parameters, instead of marginalizing over them, can give both imprecise and inaccurate estimates for other parameters of interest. Even in our toy example with simple posteriors, we find that fitted uncertainties can still be too small, by a factor of 2. In BLISS mapping, the estimated sensitivities at the knots---and so the interpolated maps---become inaccurate for the data when the photon noise is low. The maps also start fitting noise when the average data per knot is $\sim10$ or less. However, in many reasonable cases, the knot values match the intrinsic sensitivity to within the photon noise and the maps give good fits to the detector signal.

Furthermore, standard BLISS mapping is a viable shortcut to the rigorously Bayesian Jump-BLISS mapping. Both methods return similar estimates for the astrophysical model, and the knots in standard BLISS mapping behave like actual jump parameters. Curiously, our low-order polynomial model is often as precise as BLISS mapping at fitting eclipse depths, yet the latter is preferred for high-significance eclipses and more featured sensitivity variations. We also find that using the $\beta$ method to inflate uncertainties does not always increase the predictive power of fits.

In our tests, BLISS mapping does not predict the centering precision of a data set. Selecting a proper number of knots can require fine-tuning---proposed methods may not work in general. But, we find that BLISS mapping usually fits eclipse depths more accurately than precisely (i.e. conservatively), a potential benefit against low levels of extra red noise in the light curve. Overall, therefore, BLISS mapping can be an acceptable way to model \emph{Spitzer} IRAC sensitivity variations.

\section*{Acknowledgments}
The authors thank Kevin B. Stevenson (U. Chicago), Ben Farr (U. Chicago), and the anonymous referee for helpful comments that greatly improved the manuscript. JCS was funded as a Graduate Research Trainee at McGill University.

\scriptsize{
	\bibliographystyle{aasjournal}
	\bibliography{JCS_Refs_Thesis}
}

\normalsize
\appendix

\section{Choosing Parameters}
\label{sec:mod_params_explain}

\subsection{Coefficients for the Pixel Sensitivity}
\label{sec:D_sens_explain}
To pick the $c_{\ell m}$ in Equation \ref{eq:sens_vars}, we start by choosing how much all the terms added together can change the sensitivity, or $a_{v}$. We divide this value by how many coefficients we have, $n_{c}$, where we use $n_{c}=35$ because we set $n=7$. We cannot give each polynomial term the same magnitude everywhere on the pixel, so we scale the terms to be the same at some reference distance, $d_{\mathrm{ref}}$, from the pixel center. This gives us the equation:
\begin{equation}
\mathbb{C}_{\ell m} = \frac{a_{v}}{n_{c}} \left(\frac{1}{d_{\mathrm{ref}}}\right)^{\ell + m},
\end{equation}
where $\mathbb{C}_{\ell m}$ is a limit for each coefficient. By doing this, the lower-order terms will dominate inside $d_{\mathrm{ref}}$ and vice versa, so the sensitivity tends to vary more near the pixel edges. Note that the pixel centers have the highest sensitivities in the real IRAC detector \citep[e.g.][]{reach2005absolute,cowan2012thermalW}, which is not always true in this model. We decide to set $a_{v} = 0.5$ and $d_{\mathrm{ref}} = 0.1$, but other choices work, too.

We next randomly pick each $c_{\ell m} \in \left[-\mathbb{C}_{\ell m},\mathbb{C}_{\ell m}\right]$, then \emph{rescale} all these coefficients to get a chosen amplitude for the detector signal, $\Delta D$, no matter what centroids we have. For each new sensitivity map, we draw and rescale the $c_{\ell m}$ again.

\subsection{Eclipse and Phase Curve}
\label{sec:A_param_explain}
To choose the parameters for Equations \ref{eq:phase_func} and \ref{eq:A_signal}, we start with the eclipse and work backwards. We fix $t_{\mathrm{max}} = 6$ hrs and $t_{w} = 1$ hr, and because we randomly choose $t_{e} \in \left[2,4\right]$ hrs, there is always some baseline before and after the eclipse. Then we pick a value of $\Delta D_{e}$---we set $\delta_{e}$ first (to $5.0\times10^{-3}$) if we are fitting the light curve via MCMC (Section \ref{sec:synthetic_fits}) and $\Delta D$ first if making a BLISS map (Section \ref{sec:map_compare}). In the second case, the eclipse depth is about $10^{-5}$--$10^{-2}$.

Then we look at the phase model. We randomly pick $P_{\mathrm{orb}} \in \left[15,60\right]$ hrs, which gives us part of a phase curve, and $\phi_{o} \in \left[\pi\left(1-\frac{12}{P_{\mathrm{orb}}}\right),\pi\right]$, which makes sure the peak of the phase curve happens during the observation. The bottom of the eclipse should be lower in flux than the phase curve \emph{could} be, so we calculate the maximum half-amplitude, $\alpha_{\mathrm{max}}$, the phase curve could have given the other parameters. Then we randomly choose $\alpha \in \left[0.7\alpha_{\mathrm{max}},\alpha_{\mathrm{max}}\right]$, where the lower limit on $\alpha$ could be different and is just by choice.

Lastly, we pick the amount of photon noise, $\sigma$, depending on how significant we want the eclipse to be (Equation \ref{eq:signif_ecl}). Our choices give us light curves that mimic real data; other choices could work as well.

\end{document}